\documentclass[acmtog]{acmart} 
\acmSubmissionID{385}

\usepackage{booktabs}

\setcitestyle{nosort,square} %

\keywords{Geometric Deep Learning, Surface Reconstruction, Point Clouds}

\ccsdesc[500]{Computing methodologies~Point-based models}
\ccsdesc[500]{Computing methodologies~Mesh models}
\ccsdesc[500]{Computing methodologies~Neural networks}

\usepackage{hyperref}
\usepackage{amsmath}
\usepackage{graphicx}
\usepackage{comment}
\usepackage{kbordermatrix}
\usepackage{multirow,bigdelim}

\usepackage{array}
\usepackage{wrapfig}

\usepackage{csvsimple}
\usepackage{adjustbox}
\usepackage[percent]{overpic}

\definecolor{amber}{rgb}{1.0, 0.75, 0.0}
\definecolor{applegreen}{rgb}{0.55, 0.71, 0.0}
\definecolor{darkgoldenrod}{rgb}{0.72, 0.53, 0.04}
\definecolor{firebrick}{rgb}{0.7, 0.13, 0.13}

\usepackage{pgfplots}
\pgfplotsset{compat=newest}
\usepgfplotslibrary{groupplots}
\usepgfplotslibrary{dateplot}

\usepackage[bottom]{footmisc}
\newif\ifdraft
\draftfalse

\ifdraft
\newcommand{\dcc}[1]{{\color{red}[\textbf{DC:} #1]}}
\newcommand{\rhc}[1]{{\color{blue}[\textbf{Rana:} #1]}}
\newcommand{\rgc}[1]{{\color{magenta}[\textbf{Raja:} #1]}}
\newcommand{\DP}[1]{{\color{red}[\textbf{DP:} #1]}}

\newcommand{\dc}[1]{{\color{red}[\textbf{DC:} #1]}}
\newcommand{\rh}[1]{{\color{firebrick}#1}}

\newcommand{\rev}[1]{{\color{blue}#1}}

\else
\newcommand{\dcc}[1]{}
\newcommand{\rhc}[1]{}
\newcommand{\rgc}[1]{}
\newcommand{\DP}[1]{}
\newcommand{\dc}[1]{}
\newcommand{\rh}[1]{{\color{black}#1}}
\newcommand{\rev}[1]{{\color{black}#1}}
\fi

\usepackage[ruled,vlined]{algorithm2e}

\SetCommentSty{mycommfont}

\usepackage{bbold}
\usepackage{makecell}

\newcommand{\specialcell}[2][c]{%
  \begin{tabular}[#1]{@{}c@{}}#2\end{tabular}}
  
\def\hlinewd#1{%
\noalign{\ifnum0=`}\fi\hrule \@height #1 %
\futurelet\reserved@a\@xhline} 

\acmJournal{TOG}

\setcopyright{acmcopyright}\acmJournal{TOG}
\acmYear{2021}\acmVolume{40}\acmNumber{4}\acmArticle{165}\acmMonth{8} \acmDOI{10.1145/3450626.3459835}

\begin{document}

\title{Orienting Point Clouds with Dipole Propagation}

\author{Gal Metzer}
\affiliation{\institution{Tel Aviv University}}

\author{Rana Hanocka}
\affiliation{\institution{Tel Aviv University}}

\author{Denis Zorin}
\affiliation{\institution{New York University}}

\author{Raja Giryes}
\affiliation{\institution{Tel Aviv University}}

\author{Daniele Panozzo}
\affiliation{\institution{New York University}}

\author{Daniel Cohen-Or}
\affiliation{\institution{Tel Aviv University}}

\begin{abstract}
Establishing a consistent normal orientation for point clouds is a notoriously difficult problem in geometry processing, requiring attention to both \textit{local} and \textit{global} shape characteristics. The normal direction of a point is a function of the \textit{local} surface neighborhood; yet, point clouds do not disclose the full underlying surface structure. Even assuming known geodesic proximity, calculating a consistent normal orientation requires the global context. In this work, we introduce a novel approach for establishing a globally consistent normal orientation for point clouds. Our solution separates the \textit{local} and \textit{global} components into two different sub-problems. In the local phase, we train a neural network to learn a \textit{coherent} normal direction per patch (\emph{i.e.,} consistently oriented normals within a single patch). 
In the global phase, we propagate the orientation across all coherent patches using a dipole propagation. Our dipole propagation decides to orient each patch using the electric field defined by all previously orientated patches. This gives rise to a global propagation that is stable, as well as being robust to nearby surfaces, holes, sharp features and noise.

\end{abstract}

\begin{teaserfigure}
    \centering
    \includegraphics[width=\textwidth]{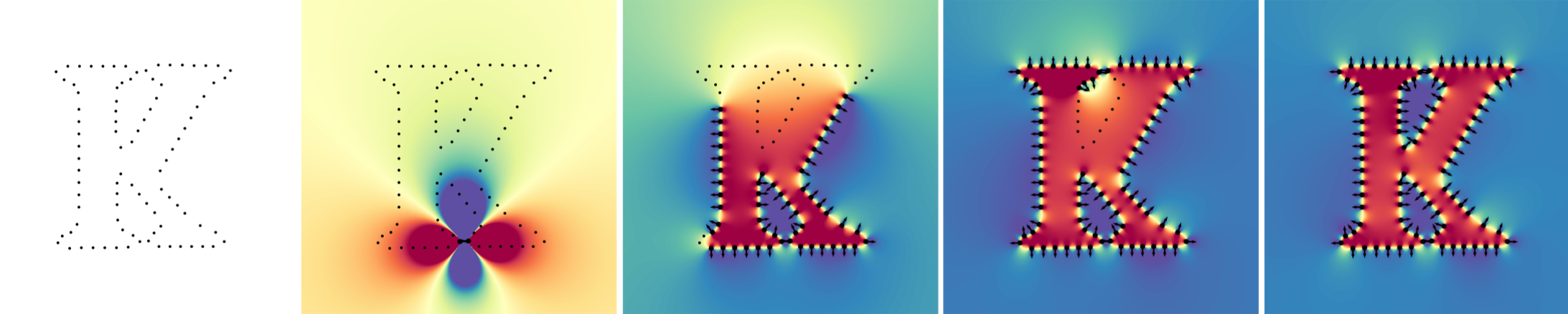} 
    \caption{An iterative dipole propagation correctly predicts a consistent normal orientation from an unstructured point cloud.}
    \label{fig:teaser}
\end{teaserfigure}

\maketitle
\section{Introduction}
Deep learning has been used to successfully synthesize point clouds for shape generation~\cite{achlioptas2018learning, li2018point, yang2019pointflow, ShapeGF}, shape completion~\cite{yuan2018pcn, wang2020point}, and up-sampling/consolidation~\cite{yu2018ec, yu2018pu, yifan2019patch, metzer2020self}. However, since standard loss functions (\emph{e.g.,} adversarial or Chamfer) do not trivially enable normal regression, these methods do not generate a globally consistent normal \textit{orientation}. 
\rh{In addition, auxiliary information needed to reconstruct a correct normal orientation (\textit{e.g.,} visibility direction) from various scanning modalities may be lost in processing (registration/re-sampling/editing) scanned point cloud data.}
Yet, a consistent normal orientation for point clouds is a pre-requisite for many techniques in computer graphics and vision, for example: surface reconstruction~\cite{kazhdan2005reconstruction, kazhdan2006poisson, kazhdan2013screened}, signing distance-fields, voxelizing volumes (\emph{i.e.,} tetrahedralization), triangulating point sets, determining inside/outside information, and rendering point sets. Thus, geometric computation is highly limited for point clouds that lack a consistent normal orientation, either struggling or failing completely to produce a meaningful solution.

Establishing a consistent normal orientation for point clouds is a notoriously difficult problem in geometry processing, requiring attention to both \textit{local} and \textit{global} attributes. The normal \textit{direction} of a point is a function of the local \textit{surface} neighborhood. Yet, since point sets do not represent the underlying surface structure, the definition of local \textit{surface} neighborhoods on point clouds is ill-defined. As such, point normals are estimated based on \textit{Euclidean} neighborhoods, which do not necessarily imply geodesic proximity (especially in regions with nearby surfaces, non-convex structures and cavities).

\begin{figure}[h]
    \centering
    \includegraphics[width=\columnwidth]{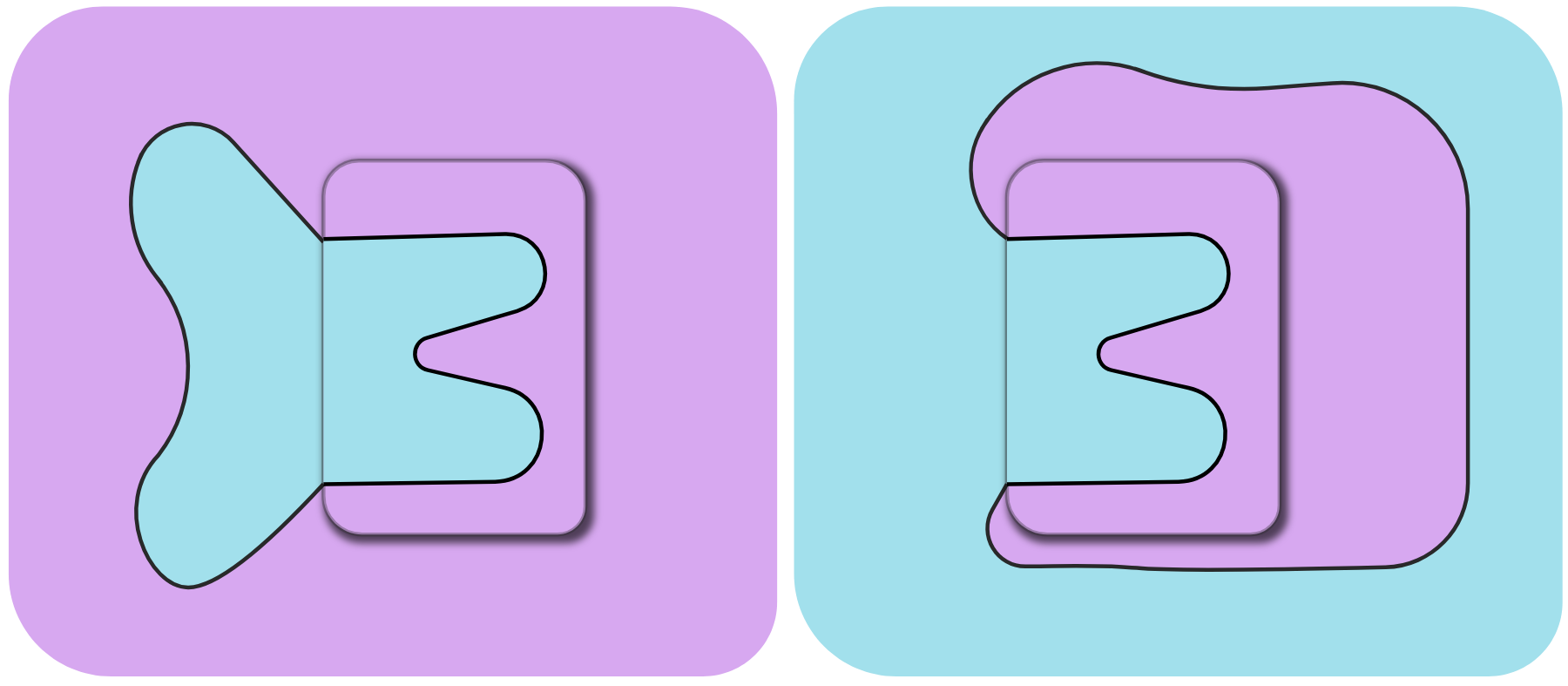} 
    \caption{The global orientation from a local patch is ill-defined. The exact same blue or purple regions in a patch can represent either inside or outside information, depending on the global context.}
    \label{fig:local_uncertain}
\end{figure}

Even when geodesic information is given, calculating a globally consistent normal orientation is non-trivial, and requires global context. Notably, when examining a flat patch from a shape in isolation, \r{establishing whether the surface plane points outward or inward is ambiguous (see Figure~\ref{fig:local_uncertain})}. Propagation techniques propose overcoming this ambiguity by starting with one correctly oriented point normal and diffusing its orientation across the entire shape~\cite{hoppe1992surface}. However, such a propagation technique assumes smoothness (\emph{i.e.,} that nearby normals are similar), leading to undesirable solutions in the case of sharp features, nearby surfaces, and noise. Moreover, the orientation accuracy is sensitive to the propagation neighborhood size; while a large neighborhood is desirable to smooth out noise and outliers, it also risks erroneously including nearby surfaces. 
A notably undesirable attribute of such techniques is their greediness, since during iterative propagation one incorrectly oriented patch degrades all subsequent propagation steps.

In this work, we introduce a new approach for orienting point clouds.
We split the orientation problem into two sub-problems, \textit{local} and \textit{global}, which are solved sequentially. We partition the point cloud into \textit{local} patches, and learn a \textit{coherent} normal direction per patch using a neural network\rh{, such that all normals are consistently pointing either inside or outside the surface}.
Specifically, we aim to \rh{detect and flip normals that do not agree with the majority direction of all input point normals in the same patch.
Note that locally there is no sense of correctness, thus we only require patch consistency in the majority direction of the input normals.
} 
Instead of propagating the \textit{global} orientation across individual points, we propagate across coherent patches. We introduce a  dipole propagation that solves global orientation by iteratively placing electric dipoles across each coherent patch. This incrementally builds a global electric field, where each new patch is oriented using the electric field of all previously oriented patches (see Figure~\ref{fig:teaser}). 

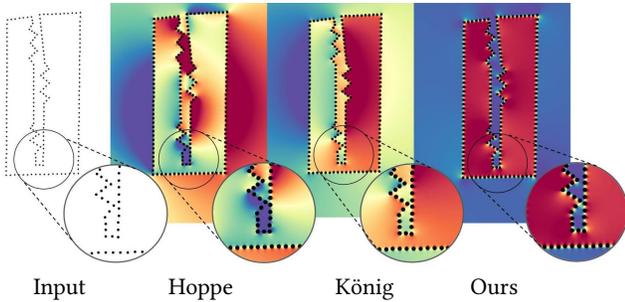
\begin{figure}[h]
    \centering
    \newcommand{\pl}{-4.3}
    \begin{overpic}[width=\columnwidth]{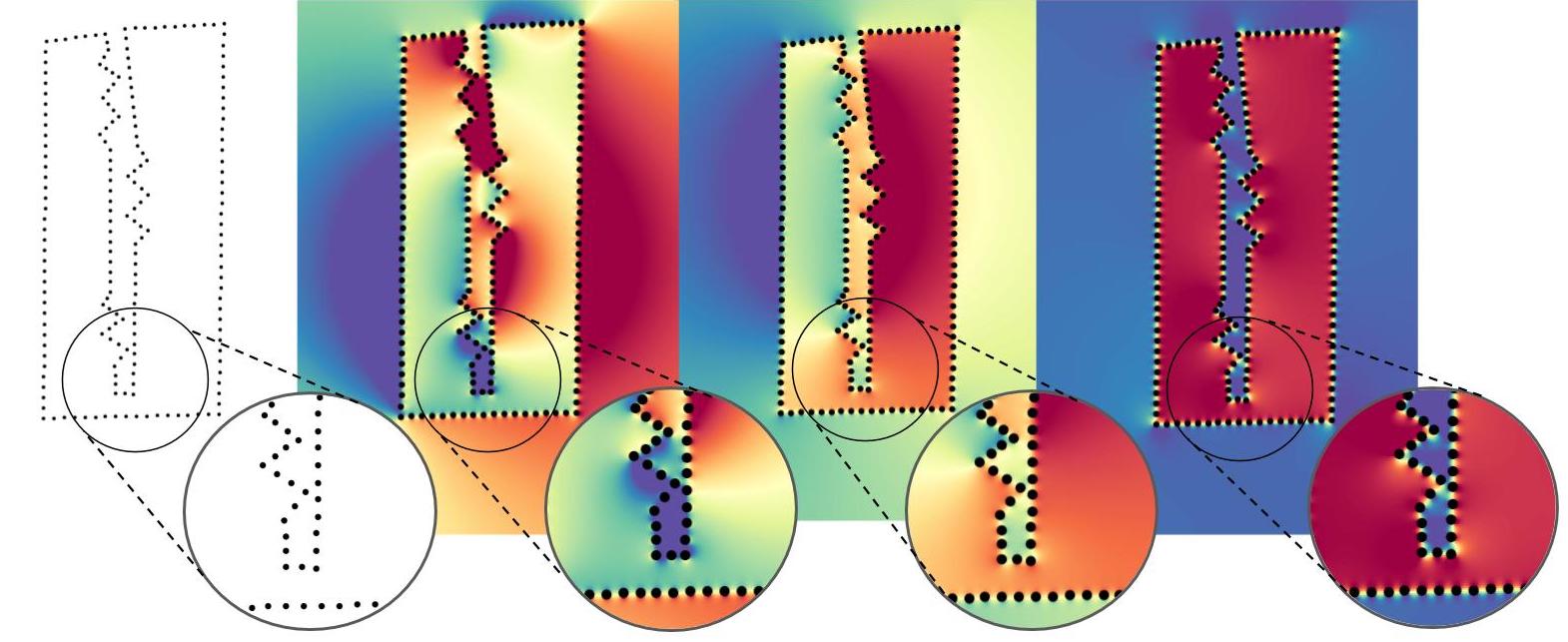}
    \put(7,  \pl){\textcolor{black}{Input}}
    \put(28,  \pl){\textcolor{black}{Hoppe}}
    \put(54,  \pl){\textcolor{black}{König}}
    \put(75,  \pl){\textcolor{black}{Ours}}
    \end{overpic}
    \caption{Starting with an input point cloud (left), the result of orientation via tangent plane normal propagation~\cite{hoppe1992surface}, or an improved version~\cite{Knig2009ConsistentPO}, both converge to an undesirable solution. Our dipole propagation is able to correctly orient each patch of coherently directed points, resulting in a desirable dipole electric potential (where blue is outside, and red is inside the surface).}
    \label{fig:baby_shark}
\end{figure}

In patches with nearby surfaces, non-convex structures, noise, and cavities, calculating a coherent normal direction is challenging and usually requires pre-defined heuristics, \emph{e.g.,} for sharp features~\cite{Knig2009ConsistentPO} or thin surfaces~\cite{xu2018towards} \rh{(see Figure~\ref{fig:baby_shark})}. Yet, this task is well suited for a neural network which can automatically learn a data-driven prior.
To this end, we train a neural network to estimate a coherent normal direction using self-supervision (points sampled from watertight meshes).
Specifically, the network learns to classify whether each input point normal \textit{agrees} with the majority direction of all input point normals in the same patch. The network-predicted probabilities are used to flip inconsistent normals within each patch, as well as guide the global dipole propagation.

Our dipole propagation calculates a consistent global orientation by iteratively propagating the correct orientation across all coherent patches using a global objective. Notably, the \textit{entire} set of previously oriented patches are used to determine the orientation of the next patch. Unlike MST-based propagation~\cite{hoppe1992surface} which only considers the previous (adjacent) orientation, our dipole propagation considers the orientation of \textit{all} previously oriented patches. Specifically, we progressively build a global electric field, which is used to orient each new patch based on the network-predicted confidence scores as well as the interaction between the electric field of all previously oriented patches. By visiting all patches (and flipping them, where appropriate), we obtain a double layer potential~\cite{folland1995introduction} whose scalar value at any point in space indicates inside or outside the shape surface. This electric potential was also referred to as the winding number for oriented point clouds~\cite{Barill:FW:2018}.

A few inconsistently oriented points only degrades the electric field locally (also noted by~\cite{Barill:FW:2018}), leading to a dipole propagation which is robust to noise and outliers. If small errors exist (\emph{e.g.,} from the neural network), they do not accumulate, which enables converging to a desirable solution in spite of the noise. Moreover, after dipole propagation is complete, we leverage the global electric field and diffuse it across all points to flip any normals incorrectly estimated by the network. \rh{Another} advantage of our dipole formulation enables building an electric field from \textit{known} normal orientations, which is useful when part of the input point cloud contains known normals (\emph{e.g.,} in point cloud upsampling).

We demonstrate the effectiveness of our technique on a variety of different 3D point clouds. We show that our neural patch coherency network can consistently orient unstructured point sets, even in the presence of noise, nearby surfaces, sparsely sampled regions, sharp features, and cavities.
In our ablation studies, we show that our neural network improves dipole propagation speed and also accuracy. Moreover, we show the applicability of our approach to clouds generated from neural networks for shape generation, shape completion, and point cloud consolidation. Our strategy also scales to large point clouds, and we show results on clouds with over one million points. In our quantitative evaluations, we demonstrate on-par or improved performance relative to state-of-the-art methods for point cloud orientation. 

\section{Related Work}
We first discuss techniques for cloud orientation, which we categorize as either surficial or volumetric. Surficial methods generally propagate the orientation information starting from a single point across the entire sampled surface. Volumetric methods aim to partition the space into inside/outside, where surface normals should point from inside to outside the surface. Then we discuss deep learning techniques  for generating point clouds (\emph{i.e.,} point locations), for up-sampling/consolidation, shape generation, and shape completion. Indeed, using such synthesized clouds for downstream tasks will also require estimating point normals, and by extension, their orientation. 

\subsection{Normal Orientation Techniques}
\subsubsection*{Surficial methods.}
The pioneering work of~\citet{hoppe1992surface}, introduced the paradigm of point normal orientation propagation via a minimum spanning tree (MST). An MST graph is built on the point cloud, where every edge between two points is assigned a weight based on the (absolute) similarity between their respective normals (later extended by~\citet{pauly2003shape} to weight distances based on an exponentially decaying function~\cite{levin2004mesh}). Starting with a single correctly oriented point normal, the orientation is propagated to nearby points across the MST graph. However, this technique is known to be highly sensitive to the choice of neighborhood size~\cite{mitra2003estimating}, since a large neighborhood irons out noise, but risks including nearby surfaces. The MST paradigm is greedy and one incorrect orientation step degrades all subsequent propagation, as we demonstrate in our comparisons (Section \ref{sec:exp}). \rh{In contrast, our work uses a global criteria to robustly handle eventual inconsistencies by considering the orientation of all previously oriented patches.}

\begin{figure}[h]
    \centering
    \newcommand{\pl}{-4.0}
    \begin{overpic}[width=\columnwidth]{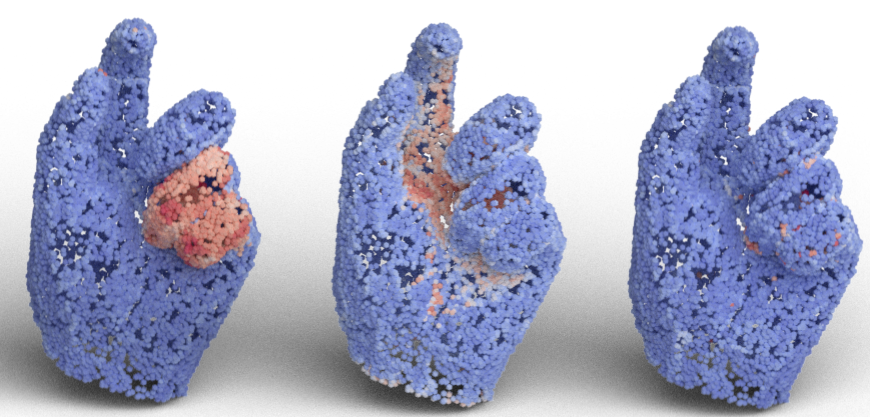}
    \put(8,  \pl){\textcolor{black}{König}}
    \put(45,  \pl){\textcolor{black}{PCP-Net}}
    \put(80,  \pl){\textcolor{black}{Ours}}
    \end{overpic}
    \caption{Normal orientation estimation, where angle errors are visualized in a heat map. The MST-based technique of~\shortcite{Knig2009ConsistentPO} fails at some point in the propagation, which results in a large region of accumulated errors. PCP-Net~\shortcite{guerrero2018pcpnet} uses only local information, which leads to a sub-optimal solution.}
    \label{fig:hands_comparison}
\end{figure}

Subsequent works proposed handling unique failure cases of the above paradigm~\cite{guennebaud2007algebraic, huang2009consolidation, xu2018towards}. For example, in the case of sharp features, \citet{xie2003piecewise} propose a multi-seed propagation that initially avoids sharp features, resulting in multiple oriented patches touching at sharp corners, which are consistently oriented in a second phase. This was later improved and extended by \citet{Knig2009ConsistentPO}, to flip normals based on the smoothness of a Hermite curve. However, our experiments demonstrate that a single mistake during propagation using \citet{Knig2009ConsistentPO} will result in a large incorrectly oriented region due to the greediness of MSTs \rh{(see Figure~\ref{fig:hands_comparison})}.

In order to mitigate errors from greedy MSTs, \citet{schertler2017towards} propose using a global optimization to fix errors, which can also handle large point clouds. Our experiments show that \citet{schertler2017towards} struggles with noise and high-frequency details, and is much slower to compute on large point clouds. For example, in Figure~\ref{fig:dragon_results} shows the favorable results of our method on a point cloud of over 1 million points which took 13 minutes to run, compared to 90 minutes for \citet{schertler2017towards}. \citet{jakob2019parallel} suggested propagating the orientation through edge collapse operations, with hand-crafted heuristics for each edge energy, as an alternative approach to MST propagation for approximating the global minimum.

To try and construct a more global criterion, \citet{seversky2011harmonic} introduces the use of harmonic functions (built from a predefined point cloud laplacian) for propagating orientation over the MST. Similar to the previous approaches, MST propagation suffers from cumulative error, and in addition the Laplacian operator for point clouds will inevitability be unstable when two surfaces are close, due to lack of geodesic connectivity, and we show an example of this problematic case in the supplementary material. We also show that our approach can robustly handle the same case.

\begin{figure}[h]
    \centering
    \newcommand{\pl}{-4.0}
    \begin{overpic}[width=\columnwidth]{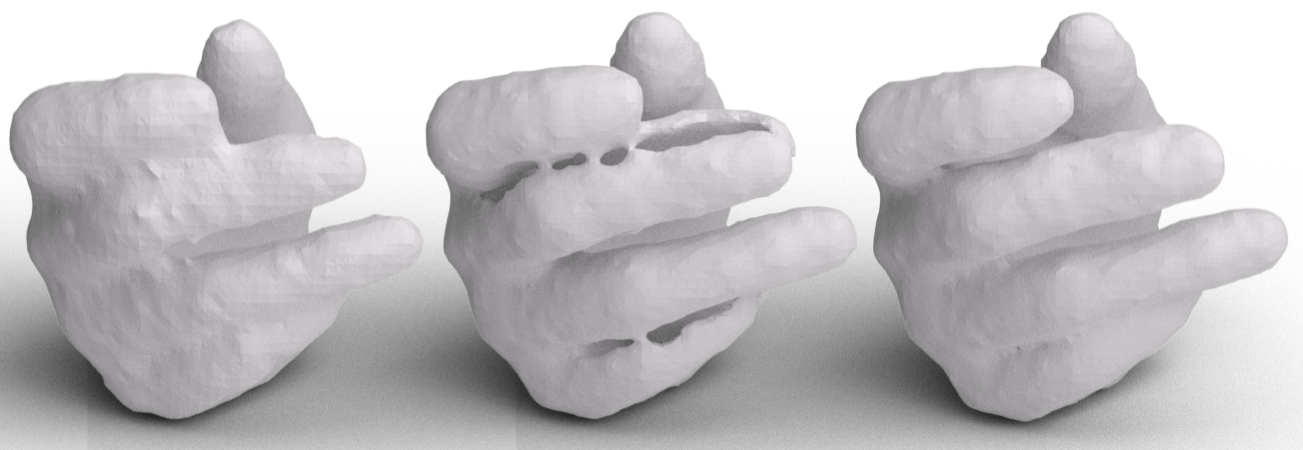}
    \put(8,  \pl){\textcolor{black}{Signing}}
    \put(45,  \pl){\textcolor{black}{König}}
    \put(80,  \pl){\textcolor{black}{Ours}}
    \end{overpic}
    \caption{Reconstruction results directly estimated by signing\rev{~\cite{mullen2010signing}}, where as the estimated normal orientation of König~\cite{Knig2009ConsistentPO} and our dipole-propagation are used as input to PSR~\cite{kazhdan2006poisson}.}
    \label{fig:hands_comparison_recon}
\end{figure}
\subsubsection*{Volumetric methods.} 
\begin{figure}[h]
    \centering
    \newcommand{\pl}{-4.0}
    \begin{overpic}[width=\columnwidth]{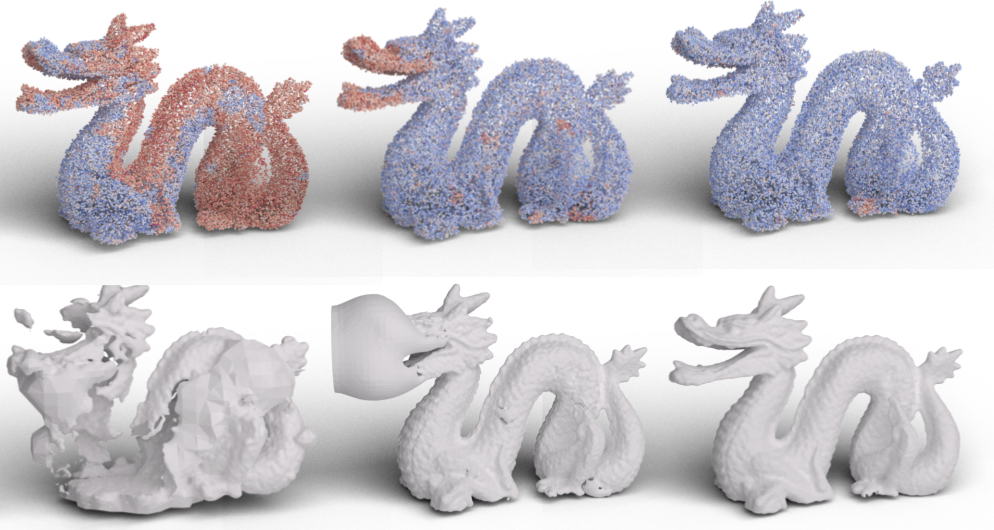}
    \put(8,  \pl){\textcolor{black}{QPBO}}
    \put(45,  \pl){\textcolor{black}{Hoppe}}
    \put(80,  \pl){\textcolor{black}{Ours}}
    \end{overpic}
    \caption{Consistent normal orientation results on a \rh{scanned} point cloud containing over 1 million points (errors visualized using a heat map). QPBO~\shortcite{schertler2017towards} contains the most error, and took 90 minutes. Hoppe~\cite{hoppe1992surface} contains less error, and took 2 minutes. Our method has fewest errors and took 13 minutes to run.}
    \label{fig:dragon_results}
\end{figure}

An alternative paradigm aims to partition the 3D space into inside/outside, where surface normals should point from inside to outside the surface. \citet{mello2003estimating} propose an adaptively subdivided tetrahedral volume to calculate the in/out function from an unoriented point set (see comparison in Figure~\ref{fig:hands_comparison_recon}, where close surfaces of the index and middle finger are incorrectly lumped together). \citet{xie2004surface} grow active contours to carve out inside/outside. \citet{chen2010binary} use binary orientation trees to determine inside/outside using point set visibility~\cite{katz2007direct}.
Some methods reconstruct a signed function from unoriented points using a variational formulation~\cite{walder2005implicit, huang2019variational}; yet they are limited to small point clouds, since their computational complexity does not scale well with resolution (see a comparison on 500 points in Figure~\ref{fig:vipss_kitten}). \rh{Whereas we demonstrate that our method scales point clouds with 10 million points.} 
Generally, volumetric methods struggle to deal with open surfaces and large holes, whereas surficial methods can struggle on sharp features and/or a greedy local objective \rh{(see Figures~\ref{fig:baby_shark}, \ref{fig:hands_comparison}, \ref{fig:dragon_results}, and \ref{fig:lion})}. 

In this work, unlike previous methods, we present a technique for point cloud orientation that leverages the power of data-driven learning with a proposed dipole propagation. Our dipole propagation defines a global objective function, and the propagation at each iteration becomes significantly less greedy, as we consider all previously oriented points. Our approach can be viewed as taking the best of the surficial approaches (operating directly on the surface points) and volumetric approaches (providing an inside / outside segmentation as a byproduct of the electric dipole field).

\subsection{Neural Point Cloud Generation}
There has been a rising interest in extending the success of deep neural networks to irregular domains. Pointnet~\cite{qi2017pointnet} pioneered the first neural network to directly consume point clouds (followed by several improved architectures~\cite{qi2017pointnet++, wang2019dynamic}), and demonstrated impressive results on discriminative tasks. This sparked interest in applying pointnet-like architectures to synthesize point clouds, for shape generation, shape completion, and up-sampling/consolidation. However, synthesizing a globally consistent normal \textit{orientation} along with each point location is non-trivial. Specifically, standard loss functions (\emph{e.g.,} adversarial or Chamfer) do not provide point-to-point correspondence on the underlying surface, which prevents defining a ground-truth normal orientation for any given synthesized point. 

\subsubsection*{Consolidation.} Deep learning techniques which generate point clouds for a downstream task (\emph{e.g.,} surface reconstruction) require some type of normal estimation / orientation on the synthesized point cloud. For example, Deep Geometric Prior (DGP)~\cite{williams2019deep} learns to consolidate point clouds, which when used in combination with Poisson reconstruction~\cite{kazhdan2006poisson, kazhdan2013screened}, results in surface reconstruction. Yet, since DGP does not regress oriented point normals, normal orientation must be solved in a post-process before using Poisson reconstruction. EC-Net~\cite{yu2018ec} consolidates point clouds, specifically on sharp edges, and requires estimating the point normals for surface reconstruction using PCA, which are indeed unoriented. Multi-step progressive upsampling (MPU)~\cite{yifan2019patch} upsample point sets using a detail-driven deep neural network, which was demonstrated to improve surface reconstruction, which also requires normal and orientation estimation. Subsequently, PU-GAN~\cite{li2019pu} proposed a generative adversarial network for upsampling point clouds, which also demonstrated improvement in surface reconstruction quality in sparse and non-uniform inputs. Self-sampling~\cite{metzer2020self} proposed consolidating point clouds using a single example, and required estimating the point normals in a post-process. Thus, employing deep learning based point consolidation / up-sampling techniques~\cite{yu2018pu}, requires normal estimation and orientation estimation for downstream tasks.
PCP-Net~\cite{guerrero2018pcpnet} proposed learning point properties (\emph{e.g.,} oriented normals and curvature) from local patches. However, since normal orientation is a global problem (\emph{i.e.,} Figure~\ref{fig:local_uncertain}), using only local information from patches leads to a sub-optimal solution (see Figure~\ref{fig:hands_comparison}).

\begin{figure*}[h!]
    \centering
    \includegraphics[width=\textwidth]{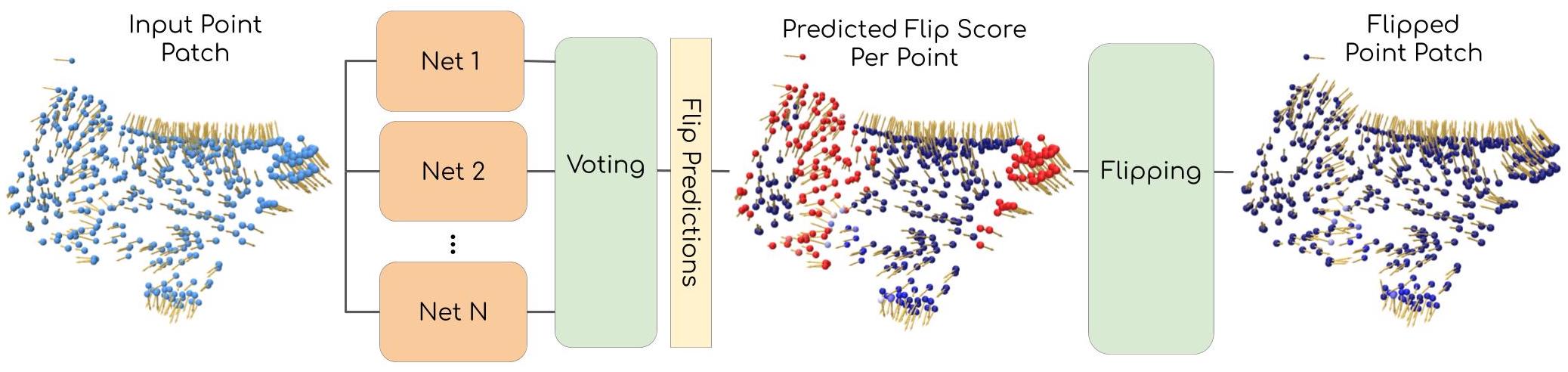}
    \caption{Coherent patch neural network. An input patch (point locations and estimated normals using Jets~\cite{cazals2005estimating}), is input to an ensemble of individual point networks which predict and vote on the probability that each point normal in the patch should be flipped. Point probabilities colored using a heat map, where high probabilities are red. \rh{The network is trained using supervised learning, where the network is encouraged to predict a high flip probability for point normals which do not agree with the majority direction the input normals.}
    After flipping the normals according to the network-predicted probabilities, we obtain a coherent normal orientation per-patch.
    }
    \label{fig:patch_overview}
\end{figure*}

\subsubsection*{Denoising.} Many deep learning techniques have been presented for point cloud denoising~\cite{hermosilla2019total, luo2020differentiable, lang2020geometric_adv}. Indeed, using the denoised cloud requires estimating normals based on the new point locations, necessitating normal orientation. Learning-based surface splatting was presented in~\cite{yifan2019differentiable}, and when denoising clouds the normals were computed using~\cite{hoppe1992surface}. PointCleanNet~\cite{rakotosaona2020pointcleannet} trained a neural network to map noisy points to clean ones and then demonstrated surface reconstruction using normals estimated using Meshlab. \citet{huang2020non} proposed non-local part-aware deep neural networks for denoising point locations, and demonstrated applicability to surface reconstruction. \citet{pistilli2020learning} developed a graph-convolutional neural network for denoising point clouds, which improved \textit{unoriented} normal estimation using off-the-shelf tools.

\subsubsection*{Synthesis.} Deep learning has been used to synthesize point clouds in a variety of \textit{generative} tasks. For example, in shape completion, neural networks are trained to synthesize points in missing regions\cite{yuan2018pcn, gurumurthy2019high, sarmad2019rl, tchapmi2019topnet, wang2020point, wen2020point, liu2020morphing}.
Another body of works generate 3D point clouds from a single-view (\emph{i.e.,} RGB image)~\cite{fan2017point, jiang2018gal}.
A wealth of deep learning techniques for shape generation using unstructured point clouds have been proposed~\cite{achlioptas2018learning, yang2019pointflow, ShapeGF}. 
A popular technique for generating point clouds is through auto-encoders~\cite{li2018so, yang2018foldingnet, groueix2018papier, liu2019l2g, zhao20193d}. PC-GAN~\cite{li2018point} presented a technique for synthesizing point clouds using generative adversarial networks. PointGrow~\cite{sun2020pointgrow} proposed an autoregressive framework for generating each point recurrently. PointGMM~\cite{hertz2020pointgmm} predicts a mixture of Gaussians which can be sampled to obtain point locations. 

Point-based neural generation techniques have focused primarily on regressing point locations, without normals. Yet, using synthesized clouds for downstream tasks will also require estimating point normals, and by extension, their orientation. Note that AtlasNet~\cite{groueix2018papier} observed a drop in reconstruction performance when regressing point normals alongside locations, compared to only regressing point locations. This empirically demonstrates that using a Chamfer assignment for point normals does not guarantee the correct normal correspondence.

\subsubsection{Deep Surface Reconstruction}
\rev{
Another line of works propose reconstructing 3D surfaces directly from point clouds.
DeepSDF~\cite{Park_2019_CVPR} learns to generate a signed distance function (SDF) from an input point cloud. 
The zero level set of the SDF can be used to reconstruct an oriented surface, as well as produce oriented normals for the input points by back-propagating through the SDF network. 
The authors describe that while training on meshes, they choose the ground truth orientation of the normals based on cameras placed around the object, and discarded shapes where the orientation was ambiguous. Point2Mesh~\cite{Hanocka2020point} reconstructs a watertight mesh from a point cloud by optimizing a MeshCNN~\cite{Hanocka2019MeshCNN} network to deform an initial mesh to shrink wrap the input point cloud. This work can utilize normal information if available, but also demonstrated results on point sets without normal orientation.
}
\section{Overview}
Our technique consists of two parts, first a \textit{local} and then a \textit{global} component. In the \textit{local} phase, we partition the shape into local patches and train a neural network to estimate a coherent normal orientation per patch. In the second phase, we use the network confidence scores to guide a dipole propagation to globally orient normals across all coherent patches.

\subsubsection*{Coherent Neural Patch Orientation.} In the first phase, we partition the shape into \rh{non-overlapping} patches and calculate the normal for each point in the patch: using Jets~\cite{cazals2005estimating}, \rh{where the normal direction is initialized by pointing away from center of mass.} The network receives as input a list of point locations and normals, and learns to predict a \textit{flip} probability per-point. Specifically, we train an ensemble of networks to predict a value between \rh{0 ($=$ should be flipped) and 1} per-point, which vote on the final flip probability. During inference, we pass (unseen) patches to the trained network and flip each normal according to network-predicted probabilities, resulting in a coherently oriented patch. An illustration of this system is shown in Figure~\ref{fig:patch_overview}.

\begin{figure}[h]
    \centering
    \includegraphics [width=0.9\columnwidth]{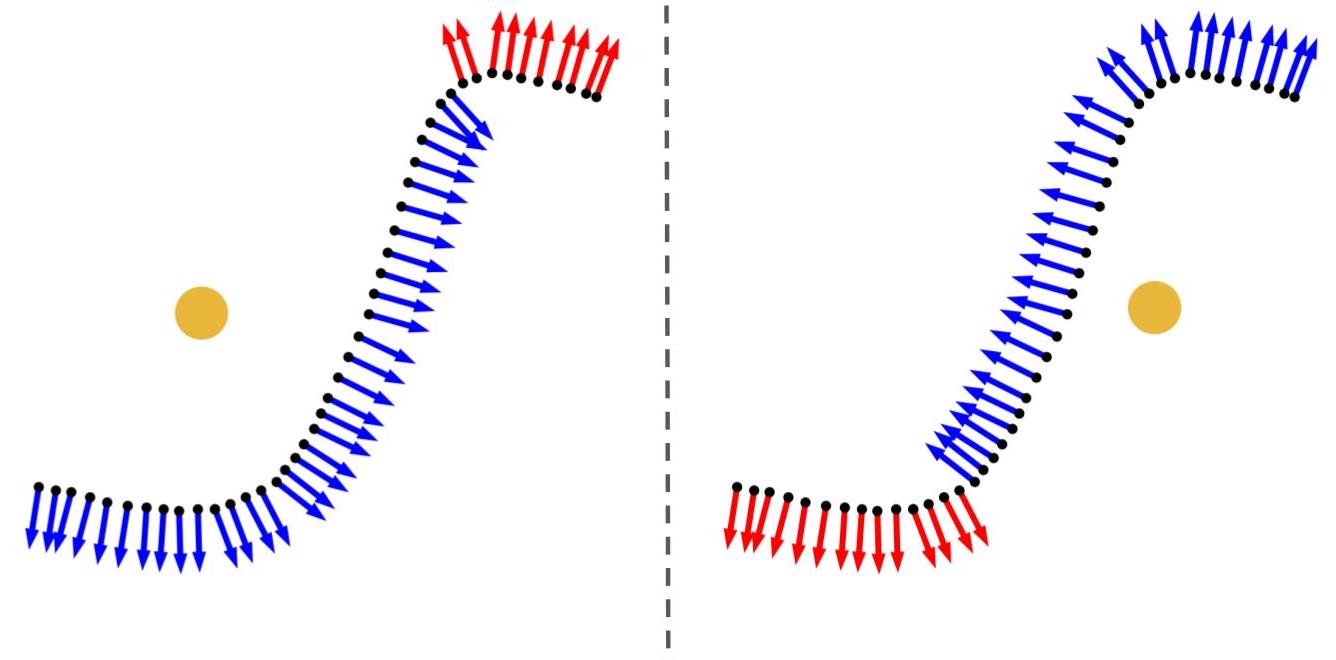}
    \caption{Different normal orientations based on the reference (yellow) point. The \textit{majority} orientation is marked in blue, which is used to create the ground-truth labels during training (red = flip).}
    \label{fig:normal_majority}
\end{figure}
\begin{figure}[h]
    \centering
    \includegraphics[width=\columnwidth]{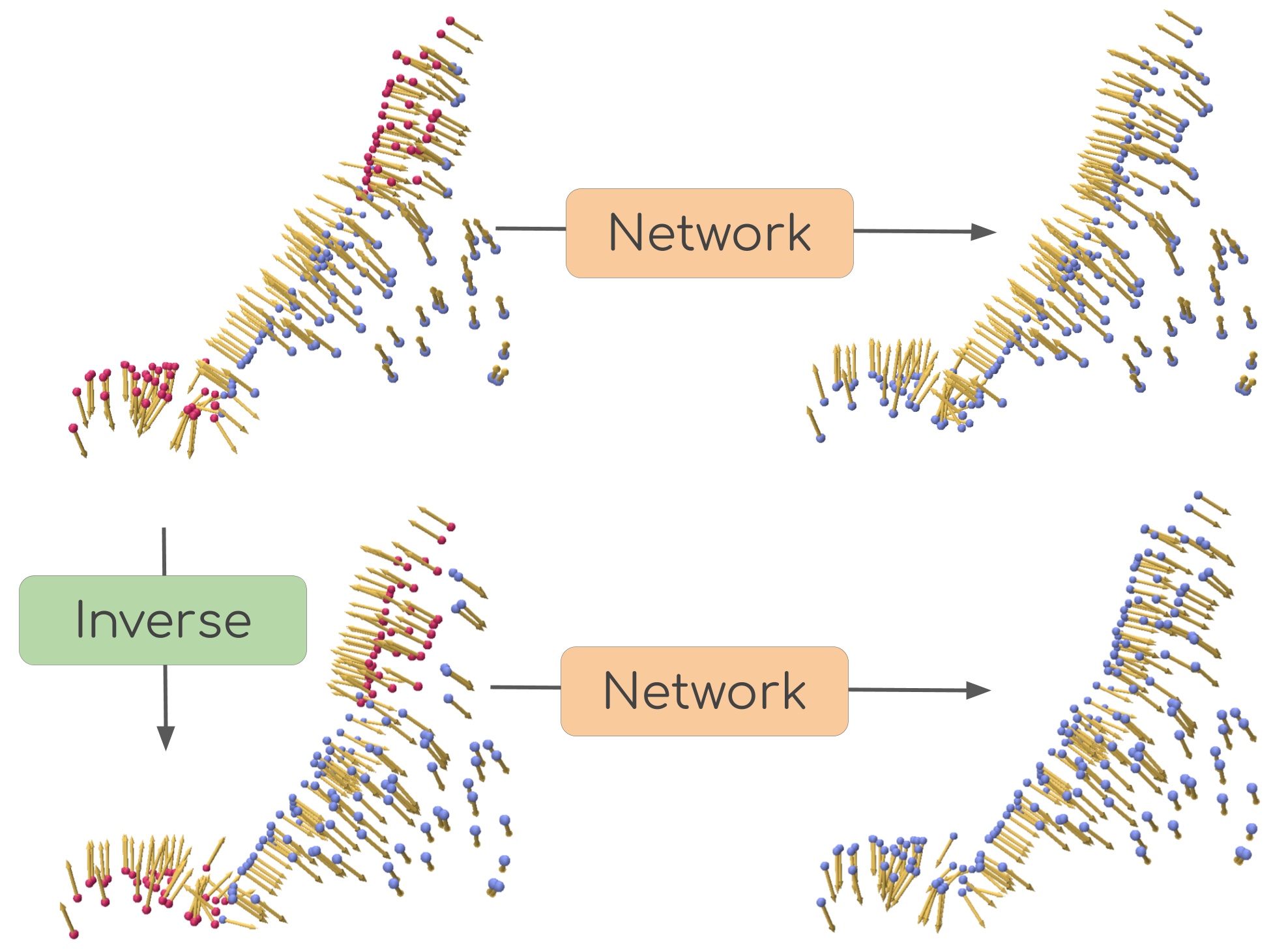} 
    \caption{The network learns to coherently orient the input patch based on the direction of the \emph{majority} of the input normals. When inverting the normal orientation on the input patch, the majority direction changes and the output orientation is similarly inverted.}
    \label{fig:coherent_patch}
\end{figure}

As mentioned previously, the global orientation is ambiguous for a single patch (see illustration in Figure~\ref{fig:local_uncertain}). As such, the network objective is to predict flip probabilities such that the point normals in the input patch are \textit{coherently} oriented, \rh{meaning that all normals are consistently pointing either inside or outside the surface.
Since training on both possibilities is a non-smooth, ambiguous objective, we prescribe the \textit{correct} coherent orientation based on the direction of the \textit{majority} of the estimated point normal directions which is used as ground-truth (see Figures~\ref{fig:normal_majority} and ~\ref{fig:coherent_patch}). The network estimated flip probabilities are also used to guide the dipole propagation to a desirable result. 
} 

Since calculating a consistent normal direction for flat or convex patches is trivial, the central challenge is estimating a consistent normal direction for patches with nearby surfaces, non-convex structures, sharp features and cavities. Yet, since these difficult cases occur less frequently \rh{we only train on patches that contain the difficult and ambiguous cases. This forces} the network to focus on the rare cases that would otherwise be ignored (due to their low prevalence in the training data).

\subsubsection*{Dipole Propagation.} 
\begin{figure}[h]
    \centering
    \newcommand{\pl}{-3.7}
    \begin{overpic}[width=\columnwidth]{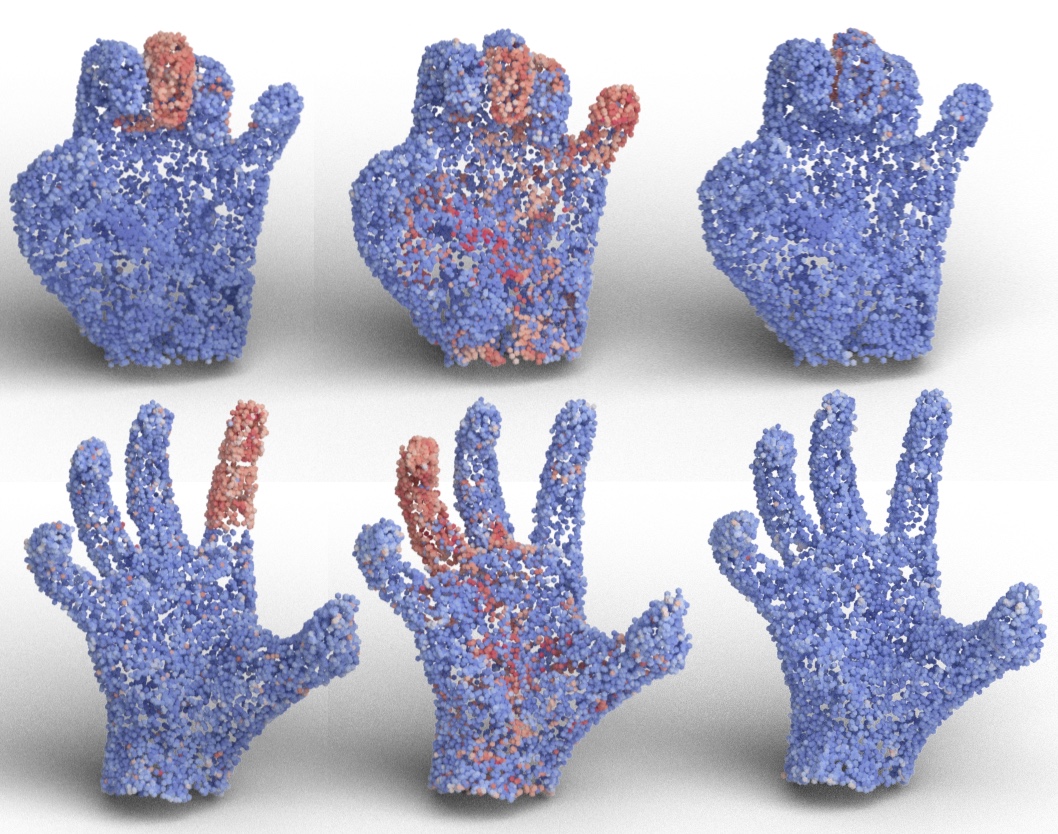}
    \put(6,  \pl){\textcolor{black}{w/o diffusion}}
    \put(39,  \pl){\textcolor{black}{w/o network}}
    \put(65,  \pl){\textcolor{black}{network + diffusion}}
    \end{overpic}
    \caption{Normal orientation results without using diffusion (left), without using a neural network (middle), and both network and diffusion (right). A desirable normal orientation result relies on both the network and the diffusion phases. The neural network not only coherently orients each patch, but also provides \textit{confidence} scores that are used during dipole propagation.}
    \label{fig:network_diffusion}
\end{figure}

\begin{figure*}[h]
    \centering
    \includegraphics[width=\textwidth]{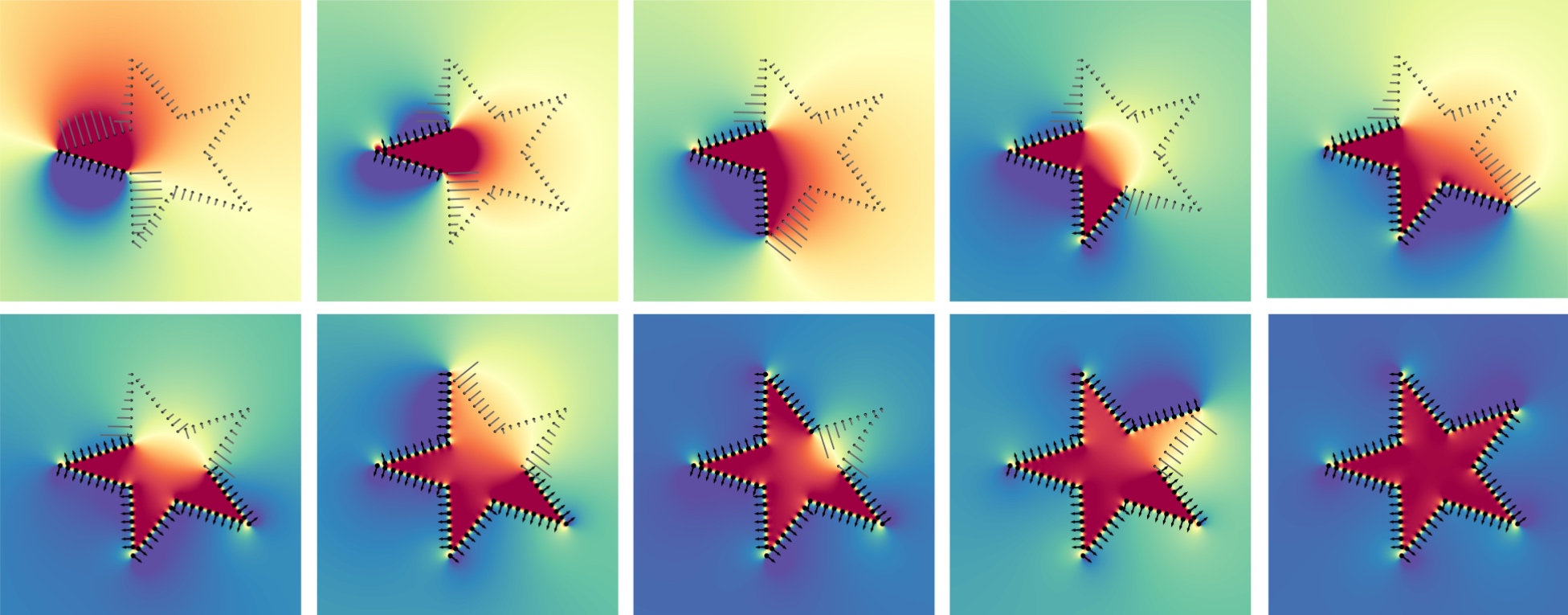}
    \caption{Iterative dipole propagation visualization. Starting with the first patch we place a dipole on every point in the patch. The electric potential dissipated by the dipoles shown in blue/red. We measure the interaction between the electric field (\emph{i.e.,} the gradient of the potential) and the normal of each point left to be oriented. The interaction is shown by a grey line with length proportional to the interaction value and direction of the unoriented normal. Note that the electric potential is only calculated for visualization purposes, and not necessary to calculate in practice.}
    \label{fig:prop_overview}
\end{figure*}

We propagate the global normal orientation to all the coherently oriented patches in the shape using dipole propagation. We start by selecting a flat patch, and treat each point in the patch as an electric dipole with polarization that points in the normal direction, which generates an electric field. Then, we find the patch with the strongest (absolute) mutual dipole interaction with the current field\rh{, weighted by the network-predicted confidences}. If the new patch interacts strongly in the opposite direction, we flip its orientation. We add the effects of the new patch to the total electric field by placing dipoles on each point in the new patch. We progressively build a global electric field, which is used to orient each new patch based on the electric field of all previously oriented patches. Once we have visited all patches (and possibly flipped them), \rh{we built an electric potential whose scalar value at any point in space indicates inside/outside the shape surface (see visualization in Figure~\ref{fig:prop_overview}). 
\rh{In the \textit{diffusion} phase, we dissipate the final electric potential to fix possible errors made by the network.}
The neural network plays a crucial role, \rh{both in terms of generating coherently oriented patches and in terms of the flip probabilities which are used to} correctly guide the propagation (see Figure~\ref{fig:network_diffusion}.)}
\rh{Note that the starting patching can have both possible orientations (all normals pointing inside or outside the surface). Upon completing propagation, we can flip the entire sign of the globally consistent normal orientation given a single correct point normal.}

\section{Method}
\subsection{Input Patch Data}
\rh{We used supervised learning to learn a mapping between input patches (point locations and normal directions), to a flip probability per point. We generate supervised pairs of input patches and ground-truth flip labels for training. Training input patches are obtained by sampling point clouds (augmented with various types of noise) from watertight meshes, and then estimating the normals for each point in the patch using an off-the-shelf tool (\emph{e.g.,} PCA or Jets~\cite{cazals2005estimating}). The corresponding ground-truth label of whether to flip each point's normal is based on the majority normal orientation direction in the patch (see Figure~\ref{fig:normal_majority}).}

\begin{figure}[h]
    \centering
    \includegraphics[width=\columnwidth]{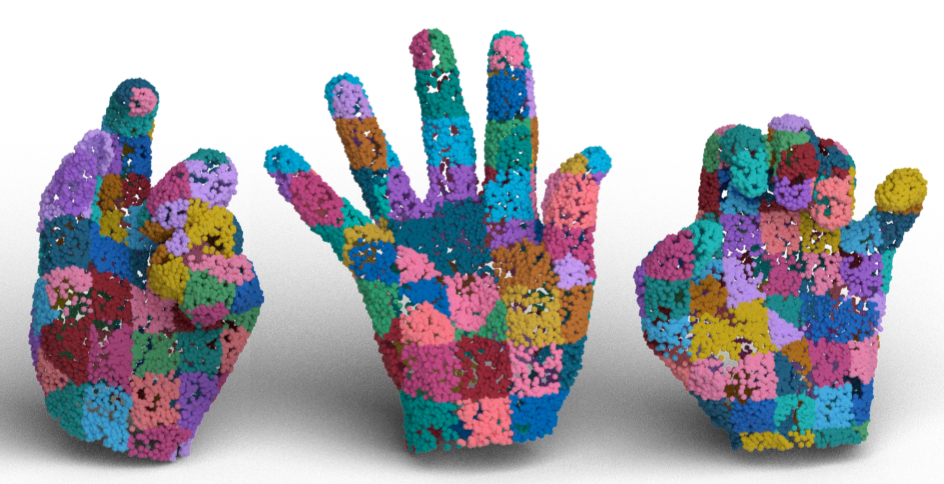}
    \caption{Visualization of our patch partitioning.}
    \label{fig:patch_visualization}
\end{figure}

Each point cloud is scaled to a unit cube and partitioned into patches corresponding to 3D cubical voxels \rh{(see Figure~\ref{fig:patch_visualization})}. After discarding empty voxel regions and merging patches smaller than $100$ points, we filter out planar patches (based on their smallest eigenvalue) since they are trivial to coherently orient (\emph{i.e.,} via a reference point on either side of the plane). We calculate normals for each point in the patch using an off-the-shelf algorithm and place each patch in a canonical \rh{axis} using PCA.

We use several forms of data augmentation to enrich the training set. During training, we flip each normal in the patch with probability $p$. This introduces uncorrelated local flip error that can be induced by noisy points, for example. We initially orient the estimated normals in each patch to point away from a single \emph{reference} point. During inference, the reference point is the center of mass. However, to induce different types of training distributions, we randomly select different reference points within the bounding box containing the patch. This augmentation generates spatially correlated flip errors. For example, in a patch that contains a corner, changing the reference point to be above or inside the corner  results in two different orientations for the perpendicular planes (see an example in Figure~\ref{fig:normal_majority}).

A strict definition of a coherently oriented patch has two valid possibilities (all normals pointing inside or outside the surface, where inside/outside cannot be defined at the patch-level). However, training on the best of the two options (\emph{i.e.,} min-distance) is non-smooth, and can oscillate between optima. In addition, this would enable the network to \textit{memorize} spatial layouts of point patches (completely ignoring the input normal information), and always predict in the $+z$ direction, for example. Therefore, during training, we define a \textit{correct} coherent orientation based on the direction of the \textit{majority} of the estimated point normal directions. Orienting point normals in the same input patch using different reference points can lead to different majority normal directions, which results in a different ground-truth label (see Figure~\ref{fig:normal_majority}).

\begin{figure}[h]
    \centering
    \newcommand{\pl}{-4.3}
    \begin{overpic}[width=0.6\columnwidth]{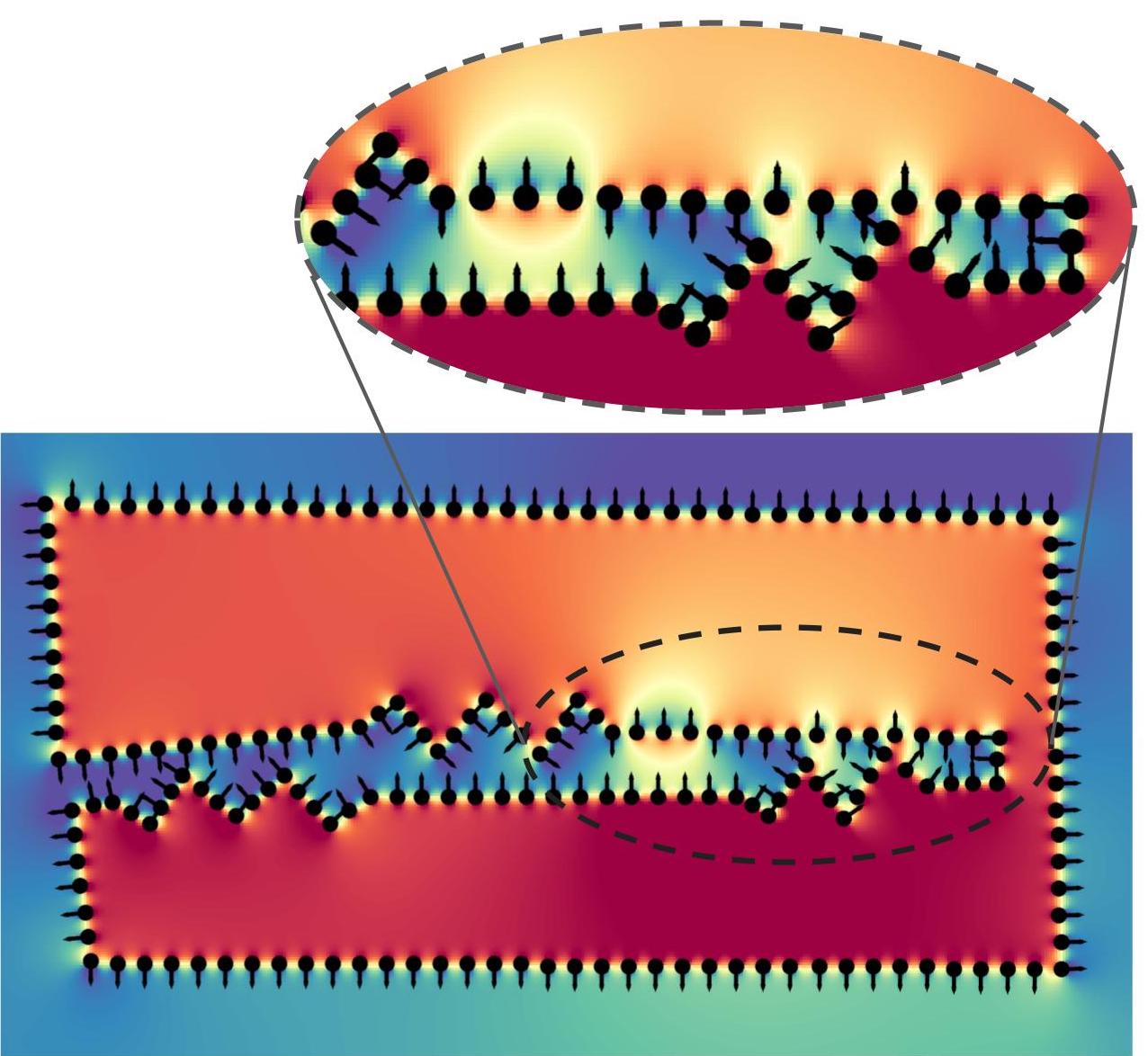}
    \end{overpic}
    \caption{Several inconsistently oriented points within a patch (5 out of 15), only degrades the electric potential locally. During dipole-propagation, if such local errors exist, \emph{e.g.,} from the neural network, they will not accumulate -- which enables converging to a desirable solution (in spite of the noise).}
    \label{fig:flip_noise}
\end{figure}

\subsection{Ensemble Training}
We train an ensemble of point neural networks to coherently orient the non-planar patches, which contain the challenging cases of cavities, non-convex, and sharp feature regions. We use a PointCNN~\cite{li2018pointcnn}-based network to predict a probability per point \rh{indicating} whether it should be flipped.
Specifically, the input to the network is a patch $\mathbb{P} \in \mathbb{R}^{N\times6}$, represented as a collection of $N$ points containing the $xyz$ positions and a normal direction for each point. The output of the network $\in \mathbb{R}^{N\times2}$ is two logits per point, which when passed to Softmax result in the $flip$ probability. The network is trained using supervision: the cross-entropy between the predicted probabilities and the ground-truth $flip$ label.

Each neural network in the ensemble can be trained with different network hyper-parameters and on different datasets. After training is complete, we average the probabilities from the different networks to obtain the final flip probability per-point. After flipping each normal according to the network-predicted probabilities the result is a coherently oriented patch. 

Learning to predict a coherent normal orientation per point requires considering \textit{both} nearby and far away points in the patch, which we address using an ensemble of multi-scale networks. The \emph{receptive-field} of the neural network naturally dictates the amount of distant points. We obtain networks with a large receptive field by using an aggressive pooling strategy. Each network is composed of PointCNN convolution layers, FPS (farthest-point sampling) pooling, and feature interpolation unpooling~\cite{qi2017pointnet++}. 
Our aggressive pooling strategy uses four FPS pooling layers, where each layer reduces the number of points by 40\%. More details about our network architecture can be found in the supplementary material.

\rh{Despite the fact that the network may produce imperfect results, our system is well equipped to handle and eventually correct such errors. For example, during dipole propagation the electric potential is robust and only degrades locally when several points are inconsistently oriented (see Figure~\ref{fig:flip_noise}). \rh{In addition, our network predicts probabilities that we use to attenuate the effects of incorrect orientations}. Finally, these lingering errors will be corrected in a second diffusion phase, which is explained in Section~\ref{sec:diffusion} (also visualized in Figure~\ref{fig:network_diffusion}).}

\subsection{Dipole Propagation}
After the network has coherently oriented each patch in the input shape, we use~\textit{dipole propagation} to calculate a consistent global normal orientation across all the patches, iteratively. We progressively build a harmonic potential \rh{(Section~\ref{sec:dipole})} from all previously oriented patches, in order to infer the orientation of new patches. Starting from a planar patch (likely to contain least amount of coherent normal errors), we treat each point in the patch as an electric dipole with polarization that points in the normal direction $\hat{n}$. The dipoles generate an electric field, which is used to orient a new patch at each iteration. To choose which new patch to orient at each iteration, we measure the absolute value of the potential energy \rh{(Section~\ref{sec:potential_energy})} at each patch which has yet to be oriented, weighted by the flip probability scores from the network. The patch with the highest potential energy is chosen for orientation, and is flipped if the energy is negative, or remains if the energy is positive. The effect of the newly oriented patch is added to the current field, and the new field (result of the previous field and with newly oriented patch), is used to orient the rest of the patches in the same iterative manner. Our decision to treat points as electric dipoles is justified in the Poisson equation \rh{(Section~\ref{sec:poisson})}, as well as the winding number~\cite{Barill:FW:2018}, which we explain in more detail below. \rh{Finally, after all the patches have been oriented by the dipole propagation, we employ an extra diffusion step to fix small errors made by the network (Section~\ref{sec:diffusion}).}
\begin{algorithm}[h]
\SetAlgoLined
\SetKwFunction{DipoleField}{DipoleField}
\SetKwFunction{FlipPoint}{FlipPoint}
\SetKwFunction{FlipPatch}{FlipPatch}
\SetKwFunction{remove}{Remove}

 $\vec{E}$ $\leftarrow$ $\mathbb{0}^{N \times 3}$ \\
 $P$ $\leftarrow$ flattest patch \\
 $remaining$ $\leftarrow$ all patches except $P$ \\

 \tcp{The current field for all points not in $P$, denoted $\overline{P}$}
 $\vec{E}$[$\overline{P}$] += \DipoleField(sources=$P$, measurements=$\overline{P}$)  \\

 \While{$remaining$ not empty}{
 \tcp{Calculate interaction (Eq.~\eqref{eq:potential_energy}) for each patch}
  i = argmax\{ {$\mid V_{j} \mid$ for j $\in$ $remaining$} \}  \\
  $P$ = $remaining$.\remove(i)  \\
  \If{$V_i$ < 0} 
  {
   \FlipPatch($P$) \tcp{flips patch if interaction negative}
   }
   \tcp{add effect of new patch to the total field}
       $\vec{E}$[$\overline{P}$] += \DipoleField(sources=$P$, measurements=$\overline{P}$)  \\
 }
 
 \tcp{Diffusion Step}
 \For{point $k \in$ point cloud} 
 {
    \tcp{calculate interaction for every point $k$ in point cloud }
    \If{$ \hat{n}_k \cdot \vec{E}[k] < 0$}
    {
        \FlipPoint($i$)  \\
    }
}
 
 \caption{Dipole Patch Propagation}
 \label{algo:patch_prop}
\end{algorithm}

\subsubsection{Electric Dipole}
\label{sec:dipole}
In electrostatics, an electric dipole arises when two oppositely charged particles are brought close together. In our scenario, each oriented point normal is treated as an electric dipole, with polarization in the direction of the normal $\hat{n}_i$. The electric potential $u$ induced by such a dipole measured at point $\vec{r}$ is
\begin{equation}
    u(\vec{r})= \frac{c_i  (\hat{n}_i\cdot \vec{r})}{4\pi  \mid \vec{r} \mid^3 },
    \label{eq:dipole_potential}
\end{equation}
where $c_i$ is the confidence score per point $i$ estimated by the neural network (proportional to the flip probability). This attenuates the effect of the points that the network is less confident about (flip probabilities close to $0.5$). The electric field is the gradient of $u$:
\begin{equation}
    \vec{E} = \vec{\nabla}u(\vec{r}) = -\frac{3c_i( \hat{n}_i\cdot \hat{r})\hat{r} - c_i \cdot \hat{n}_i}{4\pi \mid \vec{r}\mid^3}.
    \label{eq:dipole_field}
\end{equation}
Note that this is the electric field in physics up to a sign (in physics $\vec{E}=-\vec{\nabla}u(\vec{r})$). We drop the minus sign $\vec{E}=\vec{\nabla}u(\vec{r})$, since we want to orient points towards where the potential rises, rather than falls. 
Although this would imply that opposite charges repel (instead of attract), it is desirable in our orientation problem formulation. 

\subsubsection{Potential Energy}
\label{sec:potential_energy}
The potential energy between a patch and the currently dissipated field from previously oriented points is
\begin{equation}
    V_{patch} = \sum_{\hat{n_i} \in patch } c_i \cdot \hat{n_i} \cdot \vec{E}.
    \label{eq:potential_energy}
\end{equation}
The neural network confidence $c_i$ mitigates possible errors made by the network, and encourages noisy patches to be oriented last.

\subsubsection{Poisson Equation}
\label{sec:poisson}
The electric potential induced by such dipole is a solution to the Poisson equation, and has two important properties which are desirable for propagating normal orientation:
\begin{enumerate}
\item decreases with distance, and
\item negative on one side of the dipole and positive on the other.
\end{enumerate}
While an electric dipole is not the only way to obtain the above properties, its key justification lies in the Poisson equation:
\begin{equation}
\label{eq:poisson_equation}
    \nabla^2 u(\vec{r}) = 4\pi \rho(\vec{r}),
\end{equation}
which characterizes the impact of the \textit{sources} $\rho(\vec{r})$ (\emph{e.g.,} heat sources or charged particles) on the \textit{potential} $u(\vec{r})$.

In our case, the sources are the oriented points and the potential is the scalar function $u(\vec{r})$. In this way, the electric potential from all oriented points is used to orient new points.

Since Equation~\eqref{eq:poisson_equation} is linear and shift invariant, the superposition of sources $\rho_i(\vec{r})$ and solutions $u_i(\vec{r})$ is also a solution to the equations
\begin{equation}
\label{eq:superposition_equation}
    \nabla^2 u_{total}(\vec{r}) = \Sigma_i \nabla^2 u_i(\vec{r}) = \Sigma_i 4\pi \rho_i(\vec{r}) = 4\pi \rho_{total}(\vec{r}).
\end{equation}
In other words, we can obtain the total potential dissipated from all the dipole sources by summing each dipole potential (with proper coordinate shift). This leads to a closed form solution to our dipole propagation at every iteration, enabling an incremental calculation of the electric field at every time-step.
This result can also be viewed as setting the right hand side of Equation \eqref{eq:poisson_equation} to a single source, solving for Green's function, and convolving Green's function with a sum of delta functions shifted for each source.

In each step of our propagation, we use the orientation of all previously oriented points to orient new points. The oriented points are the boundary conditions, and we wish to create a smooth \textit{well-behaved} potential to decide the orientation of new points. The solution to a \textit{well-behaved} functional that satisfies minimum oscillations is the Poisson equation. Moreover, this formulation enables fast computation of the potential and its gradient in every time-step. 
Since the electric field is analytically defined at every point in space (Equation~\eqref{eq:dipole_field}), there is no need to voxelize the space and calculate the potential in every grid cell. Furthermore, Equation~\eqref{eq:superposition_equation} allows incrementally adding the effects of newly oriented points to the total potential without reevaluating it at each step of the propagation.

Finally, \citet{Barill:FW:2018} recently showed that the potential dissipated from the superposition of such dipoles is the winding number function. For shapes with oriented normals, the winding number indicates whether a point in space is inside or outside a closed surface. Indeed, after dipole propagation is complete, the final potential converges to the winding number solution.

\subsubsection{Diffusion}
\label{sec:diffusion}
After the patches have been oriented by the dipole propagation, we can employ an extra diffusion step to fix small patch coherency errors made by the network (see Figure~\ref{fig:network_diffusion}). For every patch, we evaluate the electric field of all the other oriented patches, and flip any individual points that disagree with the field. This step can be viewed as leveraging the power of the winding number function~\cite{Barill:FW:2018}, (which is stable even in the presence of a small amount of incorrectly oriented points), to correct any lingering errors. 
\begin{figure}[h]
    \centering
    \newcommand{\pl}{-4.3}
    \includegraphics [width=0.65\columnwidth]{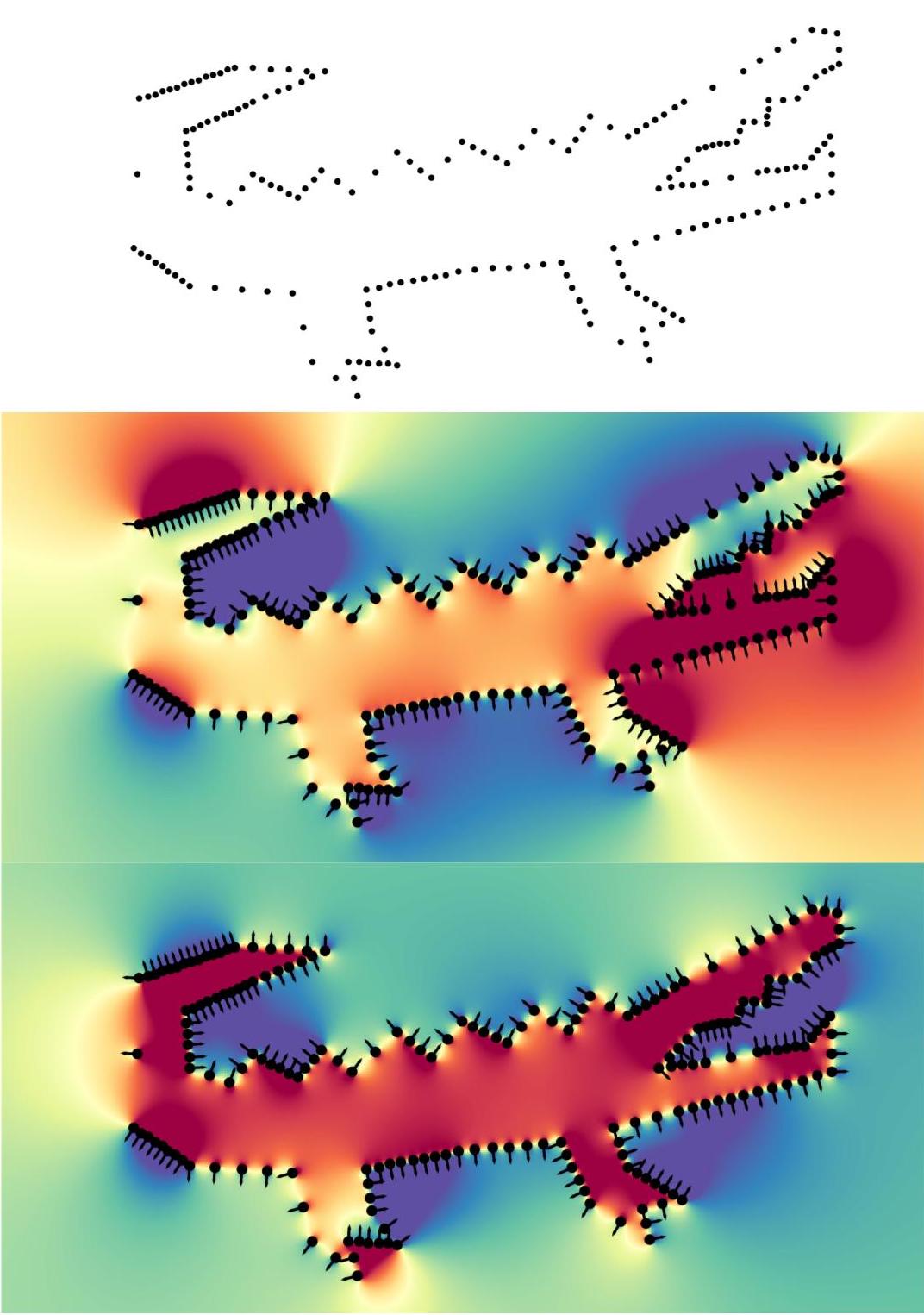}
    \caption{Top: input, middle:~\cite{Knig2009ConsistentPO}, bottom: ours.}
    \label{fig:crock}
    \vspace{-5pt}
\end{figure}
\section{Experiments}
\label{sec:exp}
We evaluate our technique on a series of qualitative and quantitative experiments on a range of different point clouds from a variety of sources. First, we explain how our method can be used for orientation interpolation in Section~\ref{sec:orientation_interpolation}. Then, in Section~\ref{sec:comparisons}, we compare to existing methods on different point cloud resolutions and neural generated point clouds, from scanning devices, and provide runtime evaluations.

\rev{
The training dataset contains 60 watertight meshes, where some are manufactured shapes (royalty free meshes collected from~\cite{Thingi10K}, ~\cite{wang2012active}), and others are non-rigid shapes from~\cite{MANO:SIGGRAPHASIA:2017}. This results in 1410 training patches, where each patch is augmented many times with noise and different reference points for the normal direction supervision. 
}

\subsection{Orientation Interpolation}
\label{sec:orientation_interpolation}
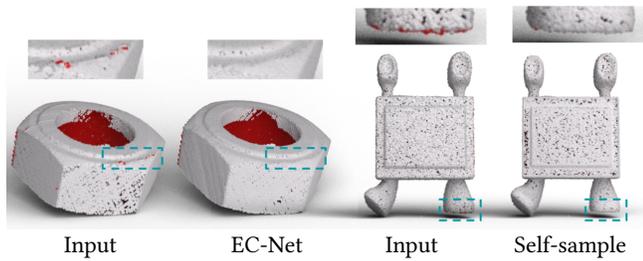
\begin{figure}[h]
    \centering
    \newcommand{\pl}{-3.7}
    \begin{overpic}[width=\columnwidth]{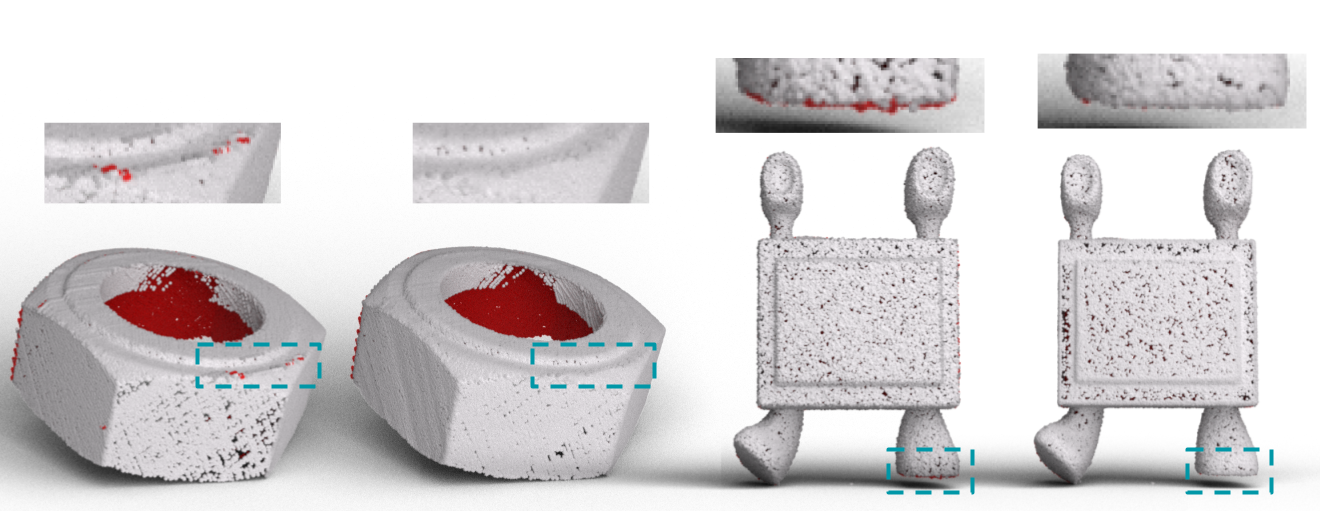}
    \put(9,  \pl){\textcolor{black}{Input}}
    \put(35,  \pl){\textcolor{black}{EC-Net}}
    \put(59,  \pl){\textcolor{black}{Input}}
    \put(79,  \pl){\textcolor{black}{Self-sample}}
    \end{overpic}
    \caption{The input point cloud contains normals (visualized using shading), which are consolidated using~\cite{yu2018ec} and~\cite{metzer2020self}. Given the input normals, we use our \textit{orientation interpolation} to predict the orientation of the newly synthesized points, resulting in accurate normal preservation. Then we enable diffusion to use the estimated electric field to re-correct the given input normals, which fixes errors (highlighted insets). Note that we (correctly) see the red (interior) since the front facing part of the surface is missing from the input point cloud scan.}
    \label{fig:orientation_interpolation}
\end{figure}

There are a couple of scenarios where it may be desirable to use a \textit{given} orientation of a set of points to decide the orientation of new points. In these scenarios, it will be convenient to build the electric field $\vec{E}$ from the given orientations, instead of propagating from scratch. For example, in the case of point upsampling we can \textit{interpolate} the newly synthesized points using the input normal orientations (see Figure~\ref{fig:orientation_interpolation}). Specifically, we calculate the field generated by the given oriented point normals in the direction of Equation~\eqref{eq:dipole_field}, and then orient the remaining points.

In the case of very large point clouds (more than $500,000$ points), we speed-up computation time by performing orientation on a sub-sampled version of the point cloud patches. Then we can orient the remaining points by calculating the field of the subset and using that to orient the remaining point normals. We use this scenario in a comparison to~\citet{schertler2017towards} in Section~\ref{sec:comparisons}, which focuses on orienting large point clouds. Our proposed speedup results in improved normal orientation accuracy and a significant computational speed up.

\subsection{Comparisons}
\label{sec:comparisons}
\subsubsection*{Low resolution point clouds}
\begin{figure}[h]
    \centering
    \newcommand{\pl}{-3.7}
    \begin{overpic}[width=0.8\columnwidth]{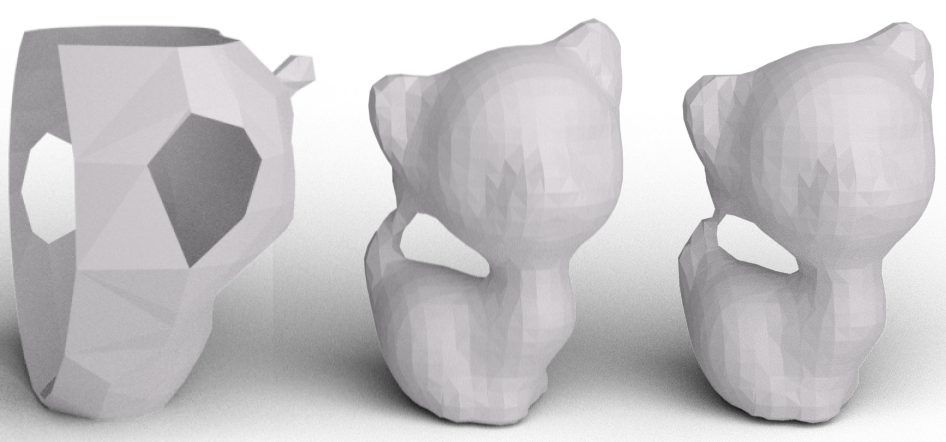}
    \put(8,  \pl){\textcolor{black}{PCP-Net}}
    \put(45,  \pl){\textcolor{black}{VIPSS}}
    \put(80,  \pl){\textcolor{black}{Ours}}
    \end{overpic}
    \caption{Surface reconstruction from a small amount of unoriented points (500) used with Poisson reconstruction~\cite{kazhdan2005reconstruction}. VIPSS~\cite{huang2019variational} is a recent variational approach which produces favorable results for small point clouds, but does not scale to point clouds above a couple thousand. PCP-Net struggles on shapes with concave parts.}
    \label{fig:vipss_kitten}
\end{figure}

In order to compare to the variational technique VIPSS~\cite{huang2019variational}, which is limited to a couple thousand point clouds, we used a point cloud from their dataset containing 500 points and ran PSR~\cite{kazhdan2006poisson} on the result of three different normal orientation methods in Figure~\ref{fig:vipss_kitten}. Since PCP-Net~\cite{guerrero2018pcpnet} predicts global orientation from local patches, it struggles on concave parts, resulting in undesirable reconstruction results. Our reconstruction result is on-par with VIPSS. However, note that their method does not scale well beyond what is presented.

\subsubsection*{Large scanned point clouds}
We demonstrate the ability of our technique to handle very large point clouds obtained from scanners~\cite{threedscans}, both in terms of accuracy and run-time. We ran a comparison on a point cloud with over one million points (Figure~\ref{fig:dragon_results}) and compared to QPBO~\cite{schertler2017towards} with \citet{xie2003piecewise} criterion, and to \citet{hoppe1992surface} (using CGAL~\cite{cgal:eb-20b} implementation). Our normal orientation (and corresponding reconstruction) obtains the best accuracy and takes 13 minutes to compute. The method of \citet{hoppe1992surface} is fastest to compute (2 minutes), but contains undesirable errors in the mouth region, resulting in poor reconstruction results. QPBO took the longest to compute (90 minutes), and contains the most error.

\begin{figure}[h]
    \centering
    \newcommand{\pl}{-3.7}
    \begin{overpic}[width=0.9\columnwidth]{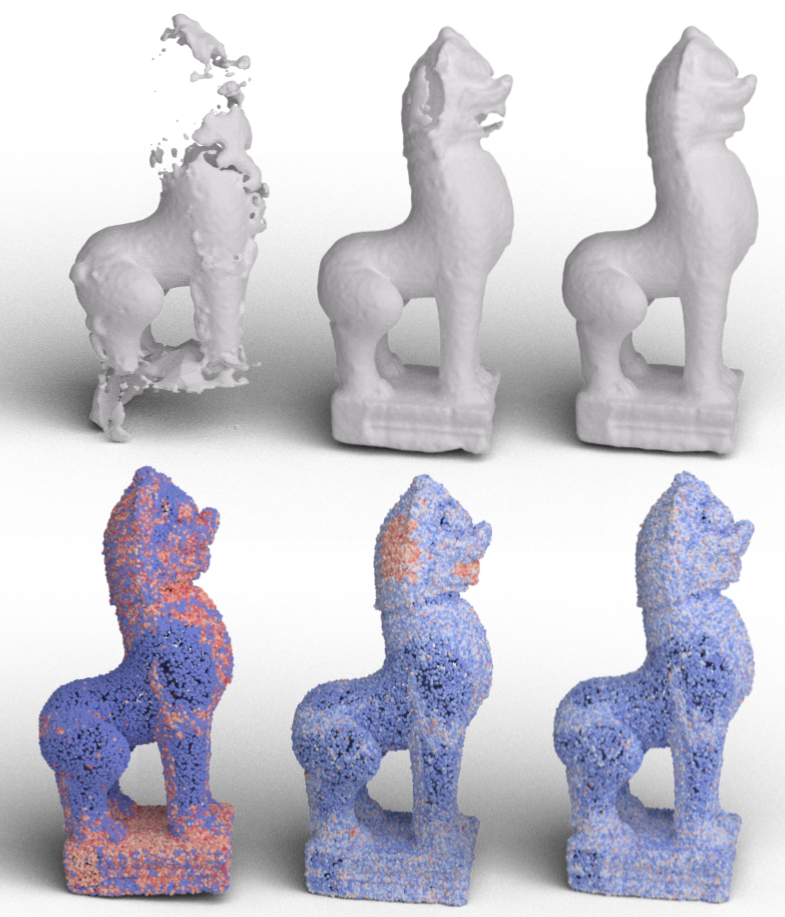}
    \put(12,  \pl){\textcolor{black}{QPBO}}
    \put(40,  \pl){\textcolor{black}{Hoppe}}
    \put(69,  \pl){\textcolor{black}{Ours}}
    \end{overpic}
    \caption{Normal orientation estimation and corresponding reconstruction results on a large (700k) \rh{scanned} point cloud.}
    \label{fig:lion}
\end{figure}

In Figure~\ref{fig:lion} we show another example of normal orientation for over 700k points with the corresponding surface reconstruction results. 
\rh{For more results on large point clouds from scanned data, including one with 10 million points, see the supplementary material. A point cloud with 10 million points takes 40 minutes to calculate normal orientation with our method (for reference, writing this point cloud to disk takes 10 minutes).}

\subsubsection*{Neural-generated point clouds}
We evaluate our method on point clouds generated using several different neural networks. Since there is no ground-truth for these examples, we visualize the estimated normal direction by shading each point with respect to the camera view (where angles larger than $100^{\circ}$ are highlighted in red). 

\begin{figure}[h]
    \centering
    \newcommand{\pl}{-2}
    \begin{overpic}[width=0.9\columnwidth]{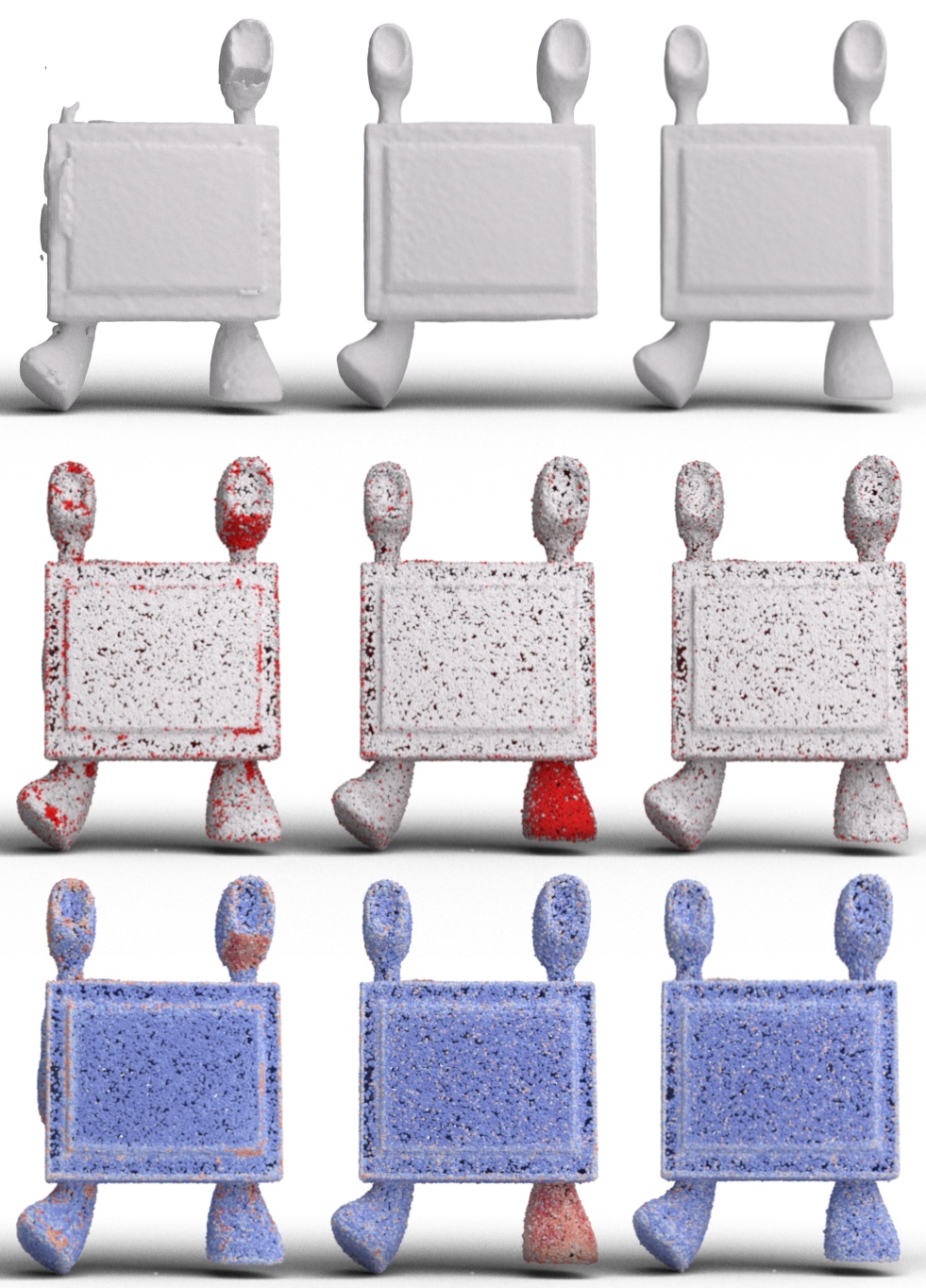}
    \put(8,  \pl){\textcolor{black}{QPBO}}
    \put(33,  \pl){\textcolor{black}{Hoppe}}
    \put(58,  \pl){\textcolor{black}{Ours}}
    \end{overpic}
    \caption{Consistent normal orientation results on a consolidated point cloud generated from~\citet{metzer2020self}. The top row contains the reconstruction results for each of the methods, and the middle and bottom rows are different visualizations (since there is no ground-truth) of the estimated normal orientations. Middle row uses a normal to viewpoint shading, and the bottom row uses a proxy to ground-truth using our interpolation propagation (shown in Figure~\ref{fig:orientation_interpolation}). The result of our method is by orientating the point cloud from scratch (without using the input normals as a reference).}
    \label{fig:self_sample_alien}
\end{figure}

\begin{figure}[h]
    \centering
    \newcommand{\pl}{-4.0}
    \begin{overpic}[width=\columnwidth]{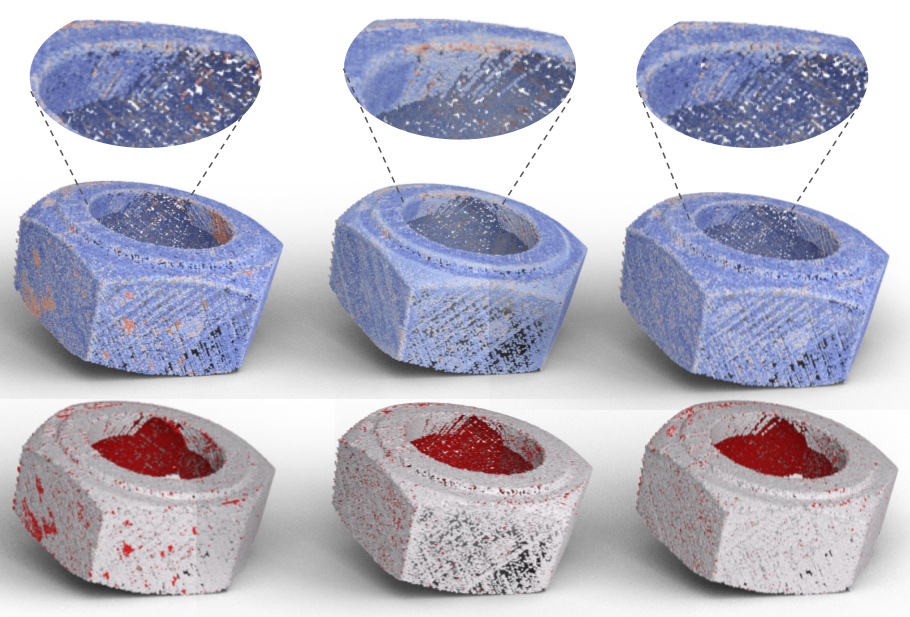}
    \put(8,  \pl){\textcolor{black}{Hoppe}}
    \put(45,  \pl){\textcolor{black}{QPBO}}
    \put(80,  \pl){\textcolor{black}{Ours}}
    \end{overpic}
    \caption{Consistent normal orientation results on a scanned point cloud consolidated using EC-Net~\cite{yu2018ec}. Since there is no ground-truth, we generated a proxy to ground-truth using the input point cloud normals and our interpolation propagation (top), as well as with thresholded viewpoint shading (bottom). The result of our method is by orientating the point cloud from scratch (without using the input normals as a reference).}
    \label{fig:screw_ecnet}
\end{figure}

In the case of neural point cloud consolidation, we generate consolidated clouds using EC-Net~\cite{yu2018ec} and \citet{metzer2020self}, and visualize the results of our normal orientation compared against two other techniques in Figure~\ref{fig:screw_ecnet} and and~\ref{fig:self_sample_alien}, respectively.
Since these techniques upsample / generate new points alongside input points which contain oriented normals, we provide an additional visualization which uses our normal orientation interpolation (shown in Figure~\ref{fig:orientation_interpolation}) as a reference to compute the angle error colors. Our visualization is useful in the case of the consolidated output of EC-Net, which has an input with an interior surface sheet which has an ambiguous orientation (see Figure~\ref{fig:orientation_interpolation}).

\begin{figure}[h]
    \centering
    \newcommand{\pl}{0}
    \includegraphics[width=\columnwidth]{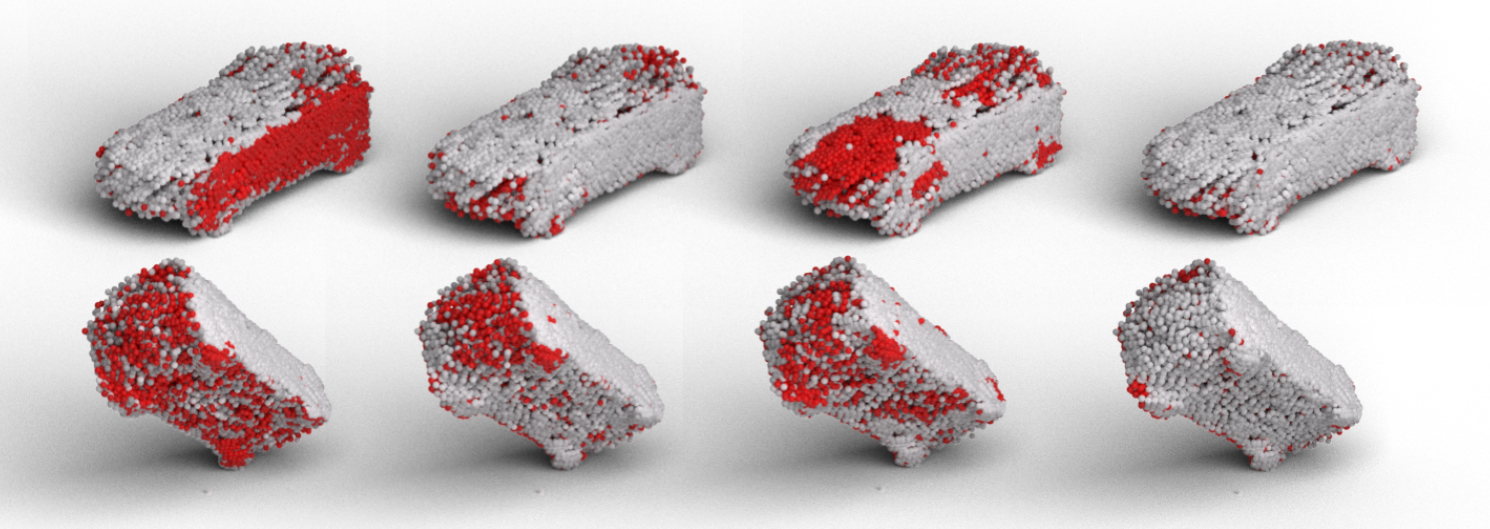} 
    \begin{overpic}[width=\columnwidth]{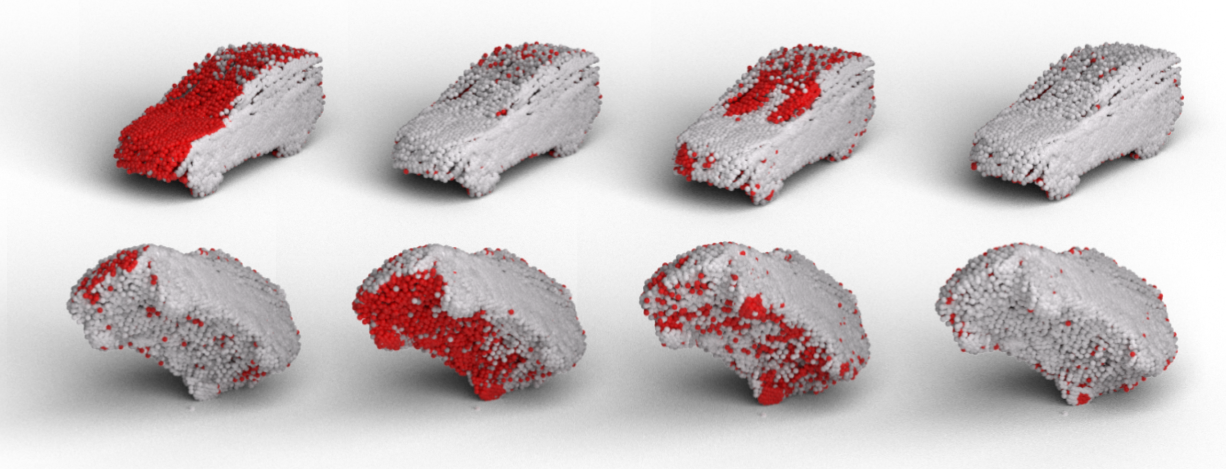}
    \put(12,  \pl){\textcolor{black}{Hoppe}}
    \put(36,  \pl){\textcolor{black}{König}}
    \put(59,  \pl){\textcolor{black}{QPBO}}
    \put(85,  \pl){\textcolor{black}{Ours}}
    \end{overpic}
    \caption{Normal orientation results on two different point clouds generated using a shape completion network~\cite{yuan2018pcn} (different viewpoint in top and bottom row). Since there is no ground-truth normal orientation, we visualize the estimated normal direction by shading each point. Note that there are also interior points inside the shape, making normal orientation particularly challenging and even ambiguous.}
    \label{fig:pcn_75_3}
\end{figure}

We also generate two different point clouds using a point completion network~\cite{yuan2018pcn}, which synthesizes point clouds without any input/given normal information. We compare the orientation results in Figure~\ref{fig:pcn_75_3}. Note that this example is especially challenging as it also contains points inside the shape (\emph{i.e,} interior points). \rh{More orientation results on point clouds generated from scratch using neural networks~\cite{ShapeGF} can be found in the supplementary material.}

\subsubsection*{Quantitative comparison and ablation}
For quantitative evaluations, we generated a collection of non-uniformly sampled and noisy point clouds which are used as input for normal orientation. This evaluation data contain two groups: (i) hands from MANO~\cite{MANO:SIGGRAPHASIA:2017} and aliens from COSEG~\cite{sidi2011unsupervised} with 5-15K points each, and (ii) scanned cloud data from~\cite{threedscans} with 700k+ points each. We calculated the percentage of correctly oriented normals in Table~\ref{table:non_uniform_hands}. Several visual results are shown in Figures~\ref{fig:hands_comparison_recon}, \ref{fig:dragon_results}, and \ref{fig:lion}, as well as in the supplementary material.

\rh{We also ran an ablation where we do not use the network, in order to highlight the effect of the neural network and presented the results in the same table (denoted as \emph{No Net}) as well as in Figure~\ref{fig:network_diffusion}. Instead of using the neural network to estimate coherent normal directions per patch, we calculated the normal orientation of each patch using Jets~\cite{cazals2005estimating} directed towards the center of mass of each patch. Then we employ dipole propagation (without any network estimated probabilities), and then diffusion. Indeed, the performance drops significantly without a network. In addition, we ran another experiment on a point-based (instead of patch-based) variant of the no-network ablation which results in another performance drop, and does not scale well (computationally) to large point clouds (see supplementary material). }
\begin{table}[h]
\caption{Quantitative comparison. Large point clouds (700k+) in the middle two rows.
The run-times for the medium sized point clouds are: $30 \pm 2$ [seconds] for our method, 3, 4 [seconds] for Hoppe and König respectively, $2$ minutes for QPBO, and $5$ minutes for PCP. }

     \begin{tabular}{|c|c|c|c|c|c|c|} %
    \hline
{Shape} & {Hoppe} & {König} & {PCP} & {QPBO} & \specialcell{No\\Net} & {Ours} \\
\hline\hline %
09-41r & 90 & \textbf{99} & 86 & \textbf{99} & 74 & \textbf{99} \\
50-50l & 92 & \textbf{97} & 90 & 95 & 74 & 95 \\
09-39l & 89 & 95 & 81 & \textbf{96} & 75 & 95 \\
17-42l & 93 & 93 & 89 & \textbf{99} & 63 & \textbf{99} \\
37-31r & 90 & 99 & 96 & \textbf{100} & 75 & \textbf{100} \\
43-16l & 90 & 98 & 88 & \textbf{99} & 75 & \textbf{99} \\
a-198 & 78 & 90 & 95 & \textbf{98} & 62 & 97 \\
a-160 & 90 & 91 & 97 & \textbf{98} & 78 & \textbf{98} \\
a-152 & 88 & 95 & 97 & \textbf{100} & 99 & 99 \\
a-188 & 82 & 86 & 95 & 96 & 90 & \textbf{97} \\
a-158 & 91 & 91 & 96 & \textbf{98} & 77 & \textbf{98} \\
a-128 & 55 & 69 & 92 & 90 & 60 & \textbf{93} \\
a-121 & 64 & 84 & 92 & \textbf{96} & 68 & 94 \\[1ex]
\hline
lion & 84 & 78 & -- & 53 & -- & \textbf{89} \\
dragon & 52 & 50 & -- & 60 & -- & \textbf{95} \\
\hline
avg & 81 & 87 & 91 & 91 & 74 & \textbf{96} \\
std & 13.23 & 12.9 & 4.76 & 14.14 & 10.4 & \textbf{2.91} \\
[1ex] %
\hline %

\end{tabular}

\label{table:non_uniform_hands}
\end{table}

\section{Discussion and Future Work}
In this paper, we presented a novel technique for estimating a consistent normal orientation for point clouds using a data-driven neural network and dipole-propagation. Our experiments demonstrate that the proposed technique is an effective tool for point cloud normal orientation.

A notable and unique feature of the orientation problem is that it is a global problem, which unlike many other consolidation problems, such as denoising or normal estimation, necessarily requires a global solution. This suggests that a careful design is needed for the solution to scale to a large number of points. Our solution tackles this using a bottom-up approach, where at the \textit{bottom} a network solves the local high-frequency problems, and the global problem is tackled by dissipating the electric field incrementally by propagation.

Locally, at the patch level, our solution is based on a unique majority-based training technique, which works surprisingly well. Training a network to predict and understand the majority is hard, and it indeed limits the size of the local patch and its variation. On the one hand, taking the patch size to the limit would allow an end-to-end solution, but would be challenging to train and nearly impossible to generalize to unseen shapes. On the other hand, overly small patches would reduce the problem to an unstable point-level solution. We believe that majority-based problems solved by networks have more intriguing potential for additional geometric problems.

An important property of dipole propagation is that it leads to a non-local traversal. This is notable in Figure~\ref{fig:teaser} (and in the accompanying video), where the patch traversal has leaps, and non-local steps. This implicitly suggests that the solution leads to a global one. However, as we showed, there are still a small amount of sporadic erroneous orientations. We recognize that some errors are unavoidable due to significant noise, yet, in the future we plan to investigate randomization techniques. For example, we can train a number of ensemble networks, where each is trained slightly differently with varying patch sizes, and during inference we propagate the results and consult them and vote. Another direction, is to remove outlier points with unstable orientation. This can possibly be achieved by coupling the orientation problem with an implicit function for reconstruction, and normal estimation. Indeed, we believe that the orientation problem is tightly-coupled with surface reconstruction, which is another interesting avenue for future work.

We are encouraged by the global nature of the dipole field, and in the future we want to consider other harmonic functions with minimum oscillations coupled with the power of neural networks. We believe that this combination can be applied to other geometric problems, to quickly infer a solution associated with an informative confidence score. 
\begin{figure}[h]
    \centering
    \newcommand{\pl}{-3.7}
    \begin{overpic}[width=\columnwidth]{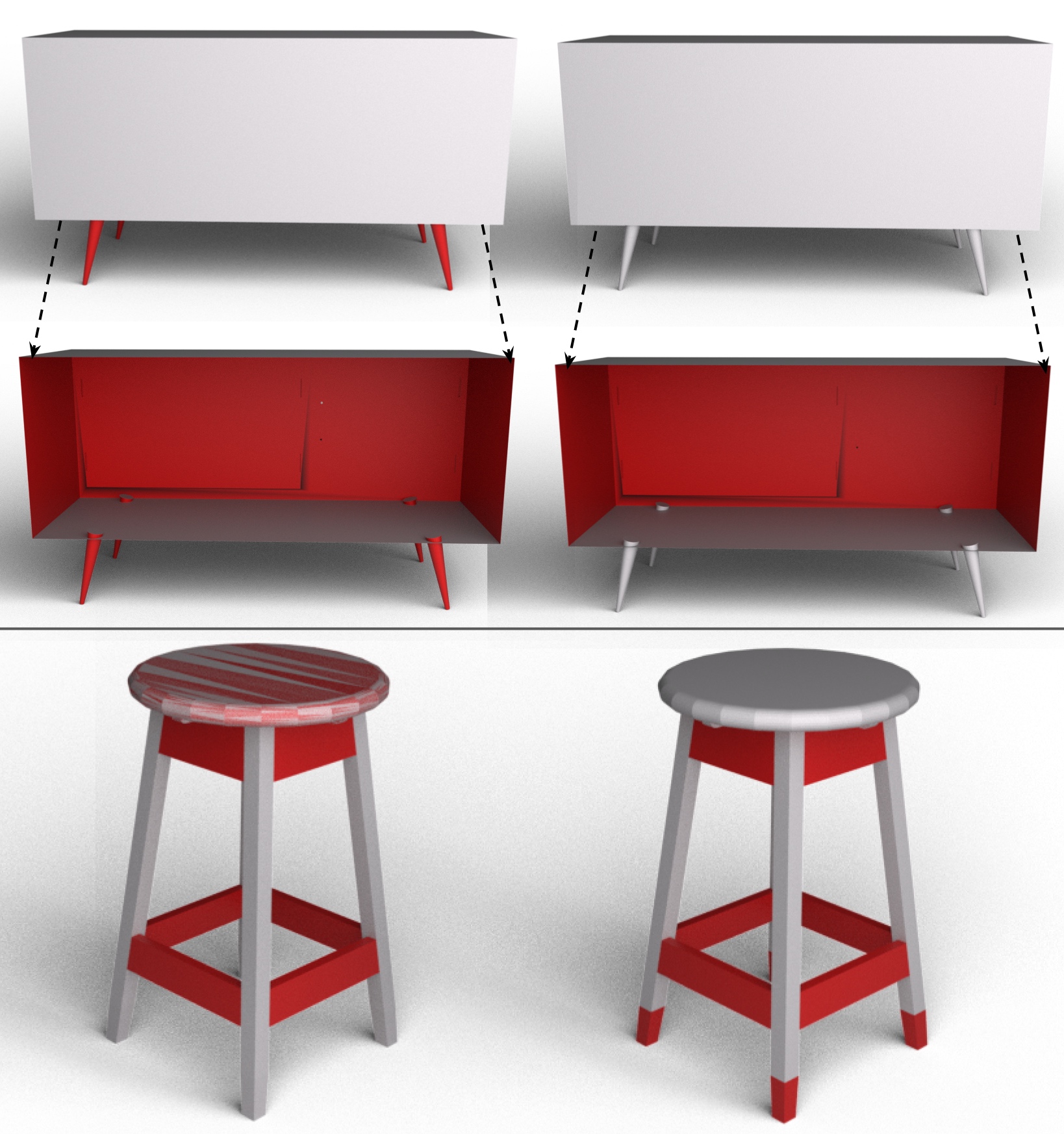}
    \put(18,  \pl){\textcolor{black}{Input}}
    \put(65,  \pl){\textcolor{black}{Output}}
    \end{overpic}
    \caption{Our technique is designed to handle point clouds which represent the exterior of an object. When using our technique to orient a general polygon soup from shapes in Modelnet, our approach works on simple models (top) with some interior protrusion from the legs, but fails to handle the challenging soup (bottom) with complex interior surfaces.}
    \label{fig:limitations}
\end{figure}
Our method is designed to orient point clouds which represent the exterior of an object. Although it may be natural to use the same approach to orient general polygon soups, for example, to obtain consistent facet normals for models in Modelnet~\cite{wu20153d}. While our approach works on simple models, it fails to handle challenging soups with multiple internal surfaces \rh{(see Figure~\ref{fig:limitations})}. Extending our approach to this case is an interesting avenue for future work.

We believe that combining dipole proportion techniques with visibility approaches has great potential in solving notoriously challenging problems in geometry processing. 

\begin{acks}
We thank Shihao Wu for his helpful suggestions, and the anonymous reviewers for their constructive comments. This work is supported by the European research council (ERC-StG 757497 PI Giryes), and the Israel Science Foundation (grants no. 2366/16 and 2492/20). This work was supported in part through the NYU IT High Performance Computing resources, services, and staff expertise. This work was partially supported by the NSF CAREER award 1652515, the NSF grants IIS-1320635, DMS-1436591, DMS-1821334, OAC-1835712, OIA-1937043, CHS-1908767, CHS-1901091, a gift from Adobe Research, a gift from nTopology, and a gift from Advanced Micro Devices, Inc.
\end{acks}

\bibliographystyle{ACM-Reference-Format}
\bibliography{bibs}

\end{document}